%% file: sensors-697681.tex
\documentclass[preprints,article,accept,moreauthors,LaTeX]{Definitions/mdpi}

\firstpage{1} 
\pubvolume{xx}
\issuenum{1}
\articlenumber{5}
\pubyear{2019}
\copyrightyear{2019}
\history{Received: date; Accepted: date; Published: date}
\pdfoutput=1

\usepackage{graphicx, import} % for pdf, bitmapped graphics files
\usepackage{float}
\usepackage{times}
\usepackage{multicol}
\usepackage{multirow}    % multi row for table
\usepackage{mathtools,amsfonts,amssymb,bm}
\usepackage{cases}
\usepackage{caption}
\usepackage{subcaption}
\usepackage[font=small]{caption}
\usepackage{colortbl}
\usepackage{xparse}
\usepackage{arydshln}
\usepackage{bm,bbm}
\newsavebox{\bigimage}
\usepackage{siunitx}
\usepackage{makecell, booktabs, caption}
\usepackage{cancel}
\usepackage{anyfontsize}
\usepackage{ifpdf}
\newcommand{\x}{\ensuremath{\times}}

\let\oldeqref\eqref
\makeatletter
\RenewDocumentCommand\eqref{s m}{%
  \IfBooleanTF#1%
  {\textup{\tagform@{\ref*{#2}}}}% If a star is seen
  {\oldeqref{#2}}%                 If no star is seen
}
\makeatother
\makeatletter
\newsavebox\myboxA
\newsavebox\myboxB
\newlength\mylenA
\newcommand*\xoverline[2][0.75]{%
   \sbox{\myboxA}{$\m@th#2$}%
   \setbox\myboxB\null% Phantom box
   \ht\myboxB=\ht\myboxA%
   \dp\myboxB=\dp\myboxA%
   \wd\myboxB=#1\wd\myboxA% Scale phantom
   \sbox\myboxB{$\m@th\overline{\copy\myboxB}$}%  Overlined phantom
   \setlength\mylenA{\the\wd\myboxA}%   calc width diff
   \addtolength\mylenA{-\the\wd\myboxB}%
   \ifdim\wd\myboxB<\wd\myboxA%
      \rlap{\hskip 0.5\mylenA\usebox\myboxB}{\usebox\myboxA}%
   \else
       \hskip -0.5\mylenA\rlap{\usebox\myboxA}{\hskip 0.5\mylenA\usebox\myboxB}
   \fi}
\makeatother
\usepackage{algorithm}
\usepackage{algorithmicx}
\usepackage{algpseudocode}
\usepackage{setspace}
\let\Algorithm\algorithm
\renewcommand\algorithm[1][]{\Algorithm[#1]\setstretch{1.5}}
\makeatletter
\def\BState{\State\hskip-\ALG@thistlm}
\makeatother
\Title{
A Novel Sensorised Insole for \\Sensing Feet Pressure Distributions%%Attention AE/ME. The following layout issues have not been checked by the English Editing Department and must be carefully verified by the AE/Layout Department: All callout issues, bold usage of callouts, and references to callouts in the text. Correct callout usage in figures. Figure and Table layout issues. Footnotes and Glossaries have not been checked. En dash usage for negative values, en dash usage to indicate relationships, en dash usage to indicate bonds (especially in chemistry). The English Editing Department is not responsible for correct italic usage for genes, proteins and technical terminology. This responsibility belongs to the authors. The following are also not checked: Number usage, ratios, en dashes for ranges, date formats, punctuation in equation lines, and less than/more than spacing (< >). Finally, capitalization and layout of titles/headings must be properly checked as well as ensuring 'Eq.' and 'Fig.' are properly spelled out, as these are layout issues.
}

\Author{{Ines Sorrentino$^{1}$*\orcidA{},
Francisco Javier Andrade Chavez$^{1}$\orcidB{},
Claudia Latella$^{1}$\orcidC{},
Luca Fiorio$^{2}$\orcidD{},
Silvio Traversaro$^{2}$\orcidE{},
Lorenzo Rapetti$^{1,3}$\orcidF{},
Yeshasvi Tirupachuri$^{1}$\orcidG{},
Nuno Guedelha$^{1}$\orcidL{},
Marco Maggiali$^{2}$\orcidH{},
Simeone Dussoni$^{2}$\orcidI{},
Giorgio Metta$^{1}$\orcidJ{}
and Daniele Pucci$^{1}$\orcidK{}}}

\address{%
${}^{1}$ \quad Dynamic Interaction Control at Istituto Italiano di Tecnologia,
  Center for Robotics and Intelligent Systems, Via San Quirico 19D, Genoa,
   Italy, franciscoJavier.andradechavez@iit.it~(F.J.A.C.); claudia.latella@iit.it~(C.L.); lorenzo.rapetti@iit.it~(L.R.); yeshasvi.tirupachuri@iit.it~(Y.T.); nuno.guedelha@iit.it~(N.G.); giorgio.metta@iit.it~(G.M.); daniele.pucci@iit.it~(D.P.)\\
${}^{2}$ \quad iCub Tech at Istituto Italiano di Tecnologia, Via San Quirico 19D, Genoa, Italy, luca.fiorio@iit.it~(L.F.); marco.maggiali@iit.it~(M.M.); simeone.dussoni@iit.it~(S.D.);\\
$^{3}$ \quad Machine Learning and Optimisation, The University of Manchester, Manchester M13 9PL, UK\\
}
\corres{Correspondence: ines.sorrentino@iit.it}

\abstract{Wearable sensors are gaining in popularity because they enable outdoor experimental monitoring. This paper presents a cost-effective sensorised insole based on a mesh of tactile capacitive sensors. Each sensor's%Confirm that your intended meaning is retained
~spatial resolution is about 4 taxels/cm$^2$ in order to have an accurate reconstruction of the contact pressure distribution. As a consequence, the insole provides information such as contact forces, moments, and centre of pressure. To retrieve this information, a calibration technique that fuses measurements from a vacuum chamber and shoes equipped with force/torque sensors is proposed.
The validation analysis shows that the best performance achieved a root mean square error (RMSE) of about $7~\si{\newton}$ for the contact forces and $2~\si{\newton\meter}$ for the contact moments when using the force/torque shoe data as ground truth.
Thus, the insole may be an alternative to force/torque sensors for certain applications, with a considerably more cost-effective and less invasive hardware.}


\begin{document}
%%%%%%%%%%%%%%%%%%%%%%%%%%%%%%%%%%%%%%%%%%
\input{sections/introduction.tex}

\input{sections/background.tex}

\input{sections/contribution.tex}

\input{sections/experiments_and_validation.tex}
\input{sections/conclusions.tex}

% %%%%%%%%%%%%%%%%%%%%%%%%%%%%%%%%%%%%%%%%%%
\vspace{6pt} 

%%%%%%%%%%%%%%%%%%%%%%%%%%%%%%%%%%%%%%%%%%
%% optional
% %\supplementary{The following are available online at \linksupplementary{s1},
%  Figure S1: title, Table S1: title, Video S1: title.}
%
% % Only for the journal Methods and Protocols:
% % If you wish to submit a video article, please do so with any other
%  supplementary material.
% % \supplementary{The following are available at \linksupplementary{s1}, Figure
%  S1: title, Table S1: title, Video S1: title. A supporting video article is
%   available at doi: link.}

%%%%%%%%%%%%%%%%%%%%%%%%%%%%%%%%%%%%%%%%%%
\authorcontributions{Resources, M.M., L.F. and S.D.; conceptualisation, I.S., F.J.A.C., S.D, N.G., M.M. and D.P; methodology, I.S.,  F.J.A.C., S.T. and D.P.; software, I.S. and F.J.A.C.; data curation, I.S., L.R., Y.T. and C.L.; validation, I.S. and F.J.A.C.; visualisation, I.S. and F.J.A.C.; writing—original draft preparation, I.S., F.J.A.C. and C.L.; writing – review and editing  I.S., F.J.A.C., N.G., L.R., Y.T., S.T., L.F. C.L. and D.P.; supervision, D.P. and G.M.}
% \authorcontributions{For research articles with several authors, a short
% paragraph specifying their individual contributions must be provided. The
% following statements should be used ``conceptualization, X.X. and Y.Y.;
% methodology, X.X.; software, X.X.; validation, X.X., Y.Y. and Z.Z.; formal
% analysis, X.X.; investigation, X.X.; resources, X.X.; data curation, X.X.;
% writing--original draft preparation, X.X.; writing--review and editing,
% X.X.; visualization, X.X.; supervision, X.X.; project administration,
% X.X.; funding acquisition, Y.Y.'', please turn to the
% \href{http://img.mdpi.org/data/contributor-role-instruction.pdf}{CRediT
% taxonomy} for the term explanation. Authorship must be limited to
% those who have contributed substantially to the work reported.}

%%%%%%%%%%%%%%%%%%%%%%%%%%%%%%%%%%%%%%%%%%
\funding{This paper is supported by EU An.Dy Project. This project has
 received funding from the European Union’s Horizon $2020$ research and
  innovation programme under grant agreement No. $731540$.}

%%%%%%%%%%%%%%%%%%%%%%%%%%%%%%%%%%%%%%%%%%
% \acknowledgments{In this section you can acknowledge any support given which
% is not covered by the author contribution or funding sections. This may
% include administrative and technical support, or donations in kind (e.g.,
% materials used for experiments).}

%%%%%%%%%%%%%%%%%%%%%%%%%%%%%%%%%%%%%%%%%%
\conflictsofinterest{The content of this publication is the sole responsibility
of the authors. The European Commission or its services cannot be held
responsible for any use that may be made of the information it contains.}

\end{document}

%% file: sections/introduction.tex
%!TEX root = ../IS_sensors2019.tex

\section{Introduction}\label{Section_introduction}

The sense of touch is pivotal in many fields of engineering, such as robotics and wearable technologies~\cite{lucarotti_synthetic_2013}. Tactile sensing allows one to characterise physical contacts in terms of contact forces and moments between bodies.
%
%In robotic applications, for example, contacts information is strategic for $i)$ controlling robot tasks, $ii)$ implementing safe human-robot interactions \cite{oneill_practical_2015}, $iii)$ implementing accurate tasks such as grasping or manipulating objects \cite{human_robot_inter}.
%
In human-centred technologies, instead, tactile sensors are embedded into wearable devices to provide feedback information on human movements. This paper contributes towards human-centred wearable technologies by proposing a novel capacity-based sensorised insole that can measure foot pressure distributions. 

A popular application of human-centred wearable  devices concerns gait analysis, that is, the systematic study of human walking. This is used in the medical field for the assessment of gait pathologies and for rehabilitation \cite{gao_fully_2016,mukhopadhyay_wearable_2015,hessert_foot_2005,sparrow_gait_2005}. Gait analysis is also used for sport purposes, that is, to help athletes improve performances while minimising the risk of injuries at the joint level \cite{queen_forefoot_2007,abdul_razak_foot_2012,orlin_plantar_2000}. Different solutions for gait event detection and plantar pressure monitoring have been proposed, including force platforms, pedobarographs, force treadmills and sensorised shoes \cite{adkin_postural_2000,cross_standing_1999,linthorne_analysis_2001,kram_force_1998}. Force platforms, pedobarographs and force treadmills are very reliable and accurate devices, but they are affected by  limitations such as high encumbrance, weight,  cost and  lack of portability, which restrict their application to laboratories.
Indeed, in some applications a high portability of the device is required and sensorised shoes are a good solution for satisfying this requirement. 
%In the state of the art, many sensorised shoes use pressure-sensitive insoles.  
%\nunosays{(}A popular application concerns the gait analysis for pathological assessment \cite{hessert_foot_2005,sparrow_gait_2005}. Instead, in sport analysis, tactile sensors are used for helping athletes to gain a high level of performance by minimising the risk of injuries .
%Several techniques and technologies to detect gait events and plantar pressure monitoring have been developed over the years\nunosays{) we go back and forth btw specifics and general view of the problem. Should focis from generic study to gait analysis, then from tctile sensing to pressure monitoring -> insoles}. 
%These solutions involve mainly the following sensors: force platforms, pedobarographs, force treadmill, sensorised shoes \cite{adkin_postural_2000,cross_standing_1999,linthorne_analysis_2001,kram_force_1998}.
%This study intends to give a contribution in this perspective. This paper presents a cost effective sensorised insole prototype with a high sensors spatial resolution providing real-time information that can be used for gait analysis. These infomation are: pressure distribution, contact forces and moments and center of pressure.
%This paper presents a novel sensorised insole prototype that has the potential to be more cost effective than existing solutions and provide information such as pressure distribution, contact forces and moments and center of pressure.

%
Recently, sensorised shoes have been gaining in popularity because they enable outdoor experimental monitoring. These devices can be equipped with accelerometers, gyroscopes, force/torque (FT) sensors and bend sensors. Thus,  sensorised shoes can measure  ground reaction forces and estimate the centre of pressure (CoP) with good accuracy. Even if such technologies are less expensive than force treadmills or force platforms, the cost is still relatively high. Furthermore, the sensors used for sensorising the shoes are heavy and rigid and do not allow subjects to walk comfortably~\cite{bamberg_gait_2008,meng_chen_intelligent_2008,schepers_ambulatory_2007,martin_schepers_ambulatory_2010}. In addition, devices such as FT sensors, accelerometers, gyroscopes, do not provide the possibility of having a distributed measurement of the pressure across the foot plant. In view of this requirement, sensorised insoles represent a good trade-off between cost and performance, although they often provide less-accurate measurements than FT sensorised shoes and typically cannot measure the horizontal ground reaction forces. However, sensorised insoles integrate  an array of tactile sensors, providing a measurement of the foot's pressure distribution. These sensors can be resistive, capacitive or piezoelectric transducers and cover at least the lateral and medial heel, metatarsal heads and the toe footprint locations. The sensors' spatial distribution allows the estimation of the pressure distribution on each foot.

%In the last years, several prototypes of insoles based on different sensing technologies have been developed \cite{zhu_umbilical_1990,abdul_razak_foot_2012,jagos_multimodal_2010}, but each of them presents weaknesses that limit their use. 

In recent years, several prototypes of insoles based on different sensing technologies have been developed~\cite{zhu_umbilical_1990,abdul_razak_foot_2012,jagos_multimodal_2010}, but several weaknesses limit their use. $i)$%are the italics necessary? Check throughout
~Some of the proposed solutions have a very low sensor spatial resolution (they use up to 10 sensors) \cite{gonzalez_ambulatory_2015, crea_wireless_2014}. Consequently, one often obtains poor information about pressure distribution, contact forces and moments. Other prototypes increase the sensor spatial resolution and even the sensors' area, but despite that, they do not cover the entire surface of the insole~\cite{benocci_wireless_2009, moticon}.  $ii)$ Some technologies are affected by drift caused by temperature changes, hardware deformation after prolonged pressure or hysteresis phenomena~\cite{gonzalez_ambulatory_2015, zhu_umbilical_1990}. $iii)$ The sensor calibration of some prototypes needs to be repeated over time, and each calibration instance is time consuming \cite{jagos_multimodal_2010, gonzalez_ambulatory_2015}. In some cases the calibration procedure is not systematic or quantitative analyses of the calibration accuracy are not performed~\cite{benocci_wireless_2009}. In addition, some of the existing solutions use a unique calibration curve for all sensors \cite{crea_wireless_2014}, which can seldom  represent the different behaviours of each sensor.

In this paper, we present a new sensorised insole prototype based on capacitive sensors that were initially conceived for the skin of the iCub humanoid robot \cite{natale2017icub}. More precisely, our main contribution is twofold. First, we present a cost-effective insole prototype that has a wide sensorised area. Second, we conceive and test a novel estimation procedure for the proposed prototype. The presented insole prototype uses capacitive transducers (i.e., taxels), grouped into modular sensors, and micro-controller tactile boards (MTBs) that drive the data acquisition \cite{cannata_embedded_2008}. The used dielectric layer is a substrate of 3D fabric using techniques adopted in the clothing industry, further reducing the cost of the sensor \cite{Maiolino2013}. The amount of individual taxels in our prototype far exceeds commercial insoles \cite{moticon,stoggl2017validation,xsensor}, providing the higher spatial resolution of 4 taxels/cm$^2$. The relationship between capacitance and estimated pressure, forces and moments is obtained through a customised calibration procedure. The calibration provides a time-effective way to obtain specific calibration models for each individual taxel. Starting from a calibration technique proposed for the same skin sensor technology in previous work \cite{kangro_skin_2017,kangro_plenum-based_2018}, we propose an improvement to overcome the load and the hardware shape limitations of our previous approaches. The new proposed method is fast and systematic, thus allowing one to find the specific calibration curve of each sensor simultaneously. Furthermore, the method calibrates the sensorised insole under pressures well beyond 1 Atm as required for usage with human subjects. Thanks to the calibration, the insole is capable of providing CoP measurements, pressure distribution over the whole foot, vertical ground reaction forces as well as the moments about horizontal axes with an accuracy comparable to commercial FT sensors. In the best case, the performances achieved by our insole, expressed in root mean square error (RMSE), were around $7~\si{\newton}$ for the contact forces and $2~\si{\newton\meter}$ for the contact moments when using force/torque sensors  as ground truth. Instead, in most dynamic cases, higher RMSE could be measured, up to $96~\si{\newton}$ for the forces and $12~\si{\newton\meter}$ for the moments.

The paper is organised as follows: Section~\ref{background} describes the technology used for building the sensorised skin of the iCub humanoid robot, the mathematical model of taxels and the methods already developed for calibrating capacitive sensors. Section~\ref{contribution} focuses on the insole prototype description and the calibration procedure. Section~\ref{experiments_and_validation} presents data extracted for the calibration experiments and describes validation experiments and results. This section also  describes the tool developed for real-time visualisation of the quantities estimated by the insole. Section~\ref{conclusions} contains a  summary of the paper and the main advantages of using our insole. It also suggests some future perspectives from the point of view of the hardware, the mathematical model and the calibration technique.

%% file: sections/background.tex
%!TEX root = ../IS_sensors2019.tex

\section{Background}\label{background}

\vspace{-0.3cm}
\subsection{Mathematical Notation}\label{subsec:math_notation}

\begin{itemize}
\item Let $\mathbb{R}$ be the set of real numbers.
\item Let $\bm x \in \mathbb{R}^n$ denote an $n$-dimensional column vector, while $x$ denotes a scalar quantity.
\item Let $||\bm x||$ be the Euclidean norm of the vector $\bm x$.
\item Let $\bm 0_{m\times n}$ be the zero matrix $\in \mathbb{R}^{m\times n}$.
\item Let $\bm I_{n}$ be the identity matrix $\in \mathbb{R}^{n\times n}$.
\item Let $f_z$, $m_x$ and $m_y$ denote the vertical ground reaction force and the moments about the horizontal axes, respectively.
\item Let $x_i \in \mathbb{R}$ denote a variable referred to the $i$-th sensor.
\end{itemize}

%%%%%%%%%%%%%%%%%%%%%%
\subsection{Tactile Sensors} \label{subsec:hardware_skin}

\begin{figure}[H]
    \centering
    \begin{subfigure}[b]{0.475\textwidth}
        \centering
        \includegraphics[width=0.8\textwidth]{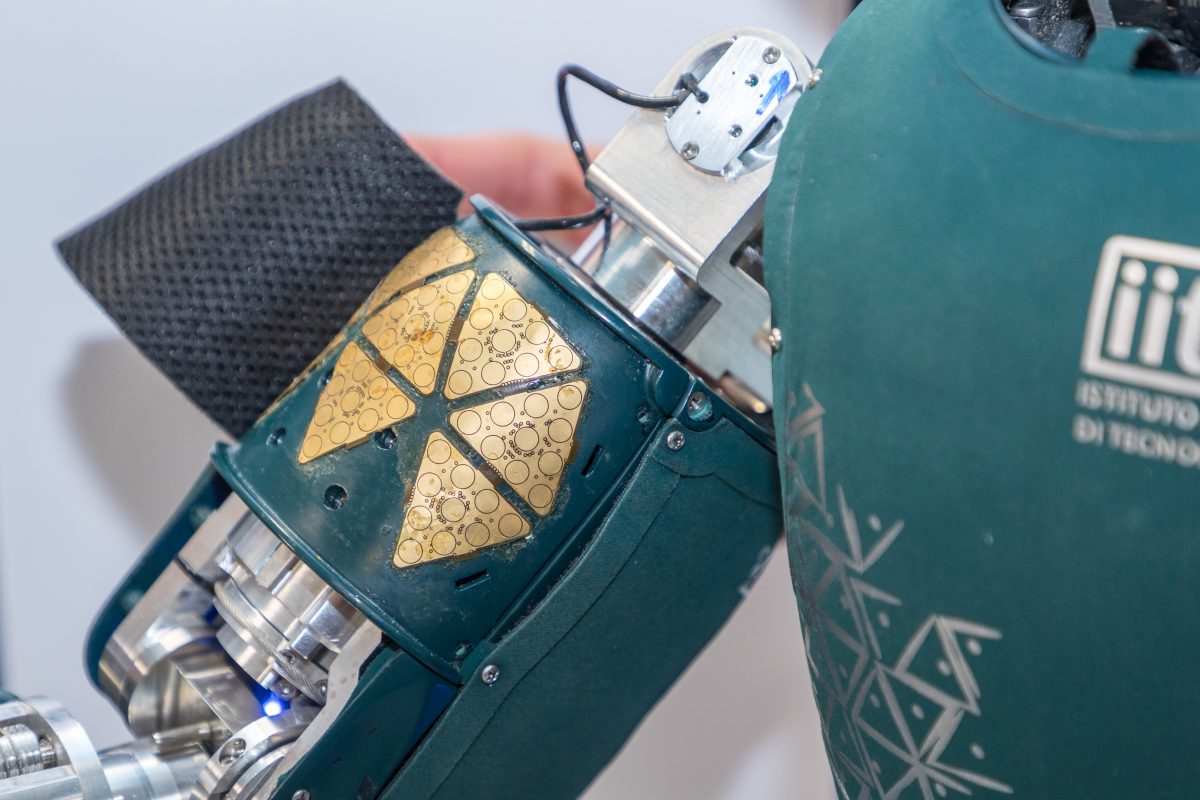}
        \caption[]%
        {{}}    
        \label{fig:iCubSkin}
    \end{subfigure}
    \quad
    \begin{subfigure}[b]{0.475\textwidth}  
        \centering 
        \includegraphics[width=0.9\textwidth]{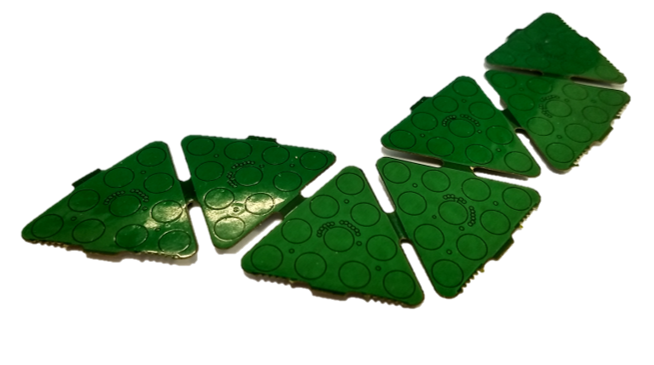}
        \caption[]%
        {{}}    
        \label{fig:triangles}
    \end{subfigure}
    \vskip\baselineskip
    \begin{subfigure}[b]{0.475\textwidth}   
        \centering 
        \includegraphics[width=0.8\textwidth]{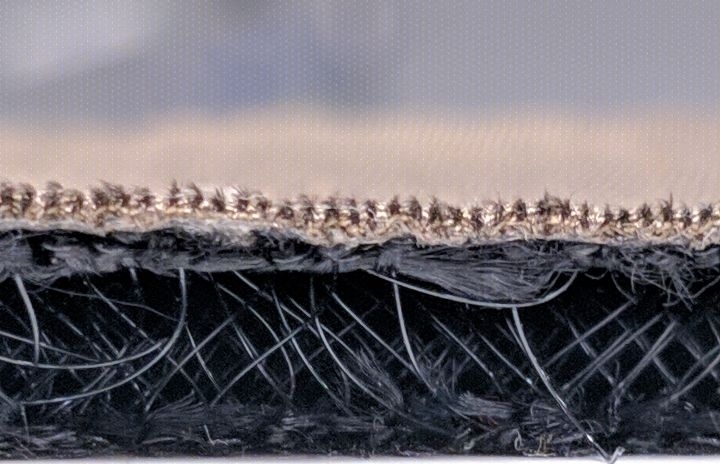}
        \caption[]%
        {{}}    
        \label{fig:dielectric}
    \end{subfigure}
    \quad
    \begin{subfigure}[b]{0.475\textwidth}   
        \centering 
        \includegraphics[width=0.5\textwidth]{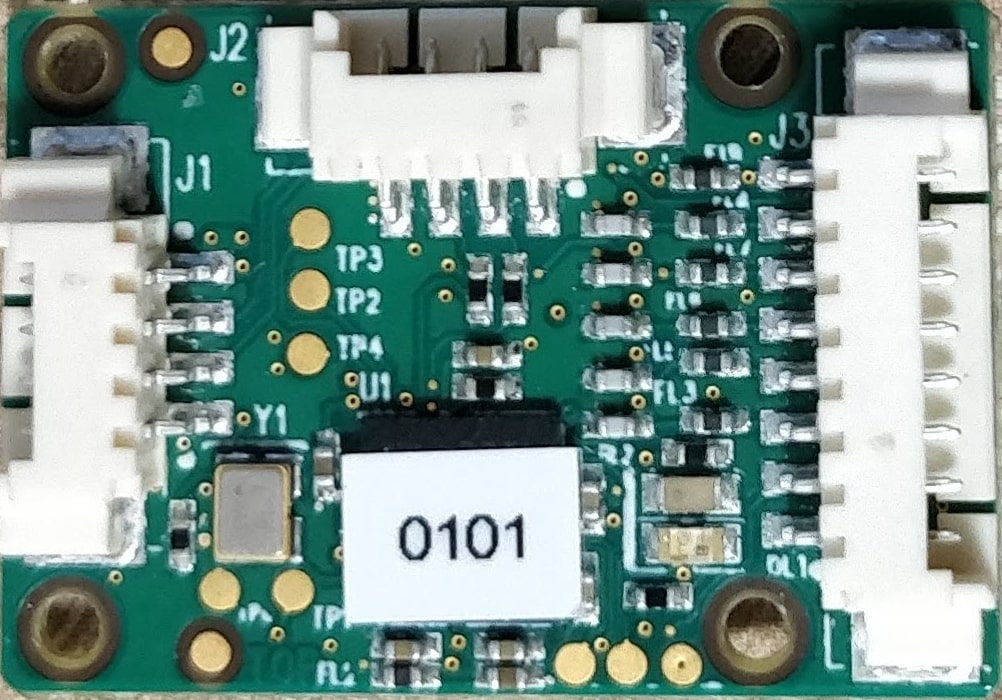}
        \caption[]%
        {{}}    
        \label{fig:MTB}
    \end{subfigure}
    \caption{({\bf{a}}) {Main components of the iCub skin.} ({\bf{b}}) Array of capacitive sensors. Each sensor implements 12 taxels and hosts the capacitive transduction electronics. ({\bf{c}}) Deformable dielectric with conductive top layer. ({\bf{d}}) Micro-controller tactile board (MTB).} 
    \label{fig:mean and std of nets}
\end{figure}

\begin{figure}[H]
 \centering
  \includegraphics[width=0.6\columnwidth]{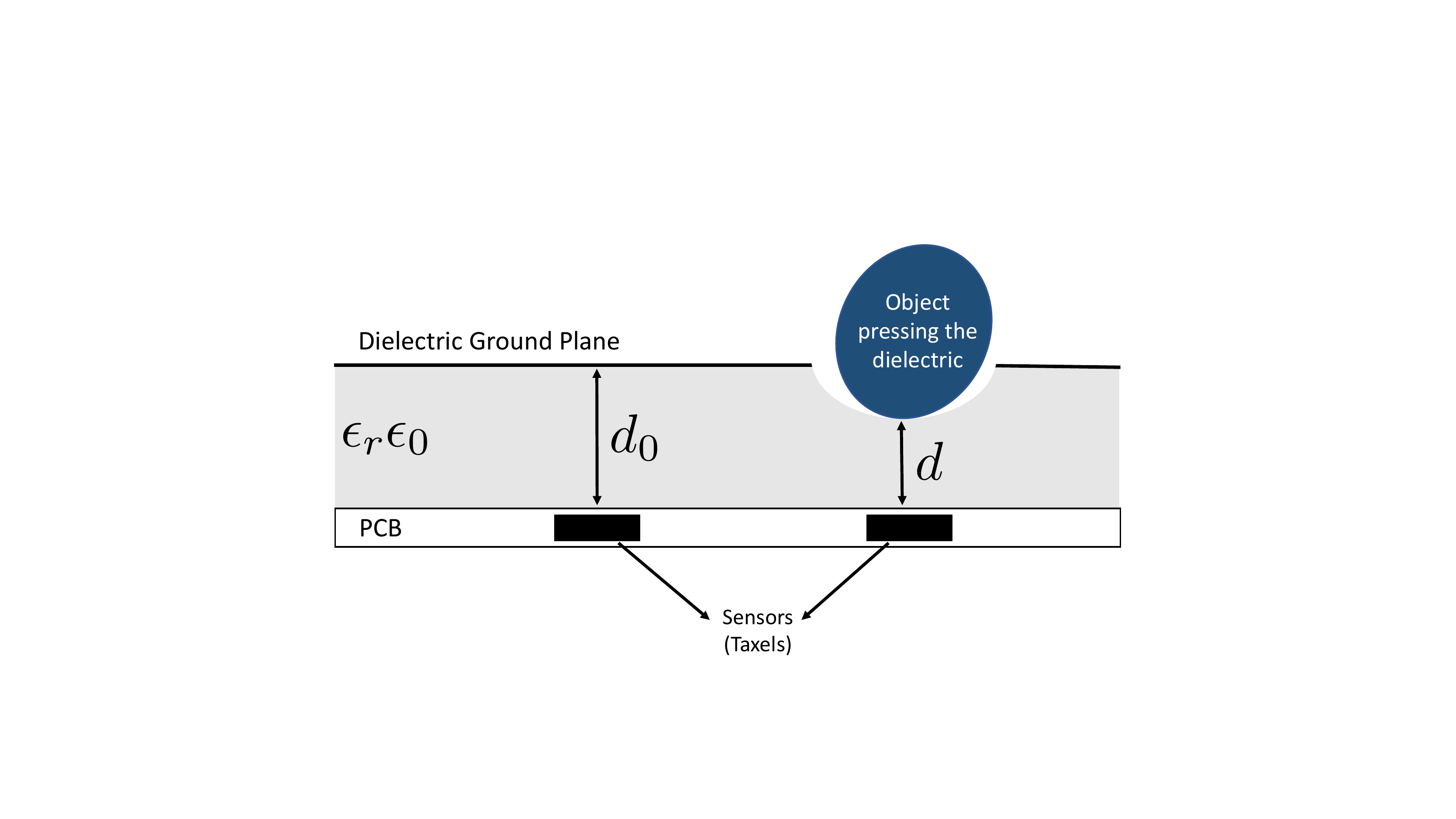}  
  \caption{ Printed circuit boards (PCBs) covered by the skin, deformed under contact pressure.}
  \label{fig:scheme_insole}
\end{figure}

The insole proposed in this paper is based on the same technology used for the skin of the iCub humanoid robot~\cite{cannata_embedded_2008}. The iCub skin (Figure \ref{fig:iCubSkin}) is an array of triangular modules of flexible printed circuit boards (PCBs), each containing 12 capacitive sensors with a circular shape (named \textit{taxels}, Figure \ref{fig:triangles}): 10 measuring the change in capacitance induced by pressure and 2 embedded in the PCB used for the temperature compensation. Each triangle has an area of 3.24 cm$^2$ and has a unique 2-bit address. The PCB provides a spatial resolution of approximately 4 taxels/cm$^2$ (the area of each taxel is approximately 19 mm$^2$). The triangles also host an integrated capacitance to digital converter (CDC) that encodes measurements of each taxel into 16-bit words with a 0.12 fF resolution. A patch of PCB is composed of many interconnected triangles where each connection includes a dedicated Inter-Integrated Circuit (I2C) bus with four synchronised serial data lines and a shared serial clock. This allows up to 16 triangles to be connected to the same MTB. The MTB is a micro-controller-based board that interfaces the I2C%Define if appropriate
~bus and the CAN%Define if appropriate
~bus (Figure~\ref{fig:MTB}). The micro-controller recognises the devices connected to the I2C bus (i.e., the CDCs), reads measurements sent by the CDC of each triangle on the bus and performs an on-board temperature and an offset compensation. Then, the MTB cuts the 16-bit words into 8-bit words suitable for communication through CAN bus. The measurements are sent to a host through the CAN bus at a rate of 50 Hz.
A deformable dielectric layer, coupled with a ground plane, is placed on top of the sensors (Figure \ref{fig:dielectric}). The dielectric layer consists of a nylon mesh with good mechanical properties in terms of hysteresis and deformation response. The mesh is provided by commercial suppliers and is cost-effective. When it is pressed, the distance between sensors and the surface of the dielectric decreases, causing a change in capacitance. That is, the higher the pressure on the dielectric, the higher the capacitance variation, as shown in Figure \ref{fig:scheme_insole}. 
The user can set a bit-shift parameter that selects which 8 bits {are considered}%was "consider", confirm that your intended meaning is retained
~of the 16. The settable value is in the range{ from} 0 (the MTB selects the least significant byte){ to}%,
~7 (the MTB selects the most significant byte). With a higher bit-shift value, the 8-bit word maps a larger range of analog capacitance values but reduces the sensibility to very small variations. The choice of this value depends on the specific application and on the specific stiffness of the used dielectric material. 


%please cahnge all subfigure in the caption into bold
%%%%%%%%%%%%%%%%%%%%%%
\subsection{A Mathematical Model for Taxels} \label{subsec:theoretical_model_taxel}

%By construction, 
The behaviour of each taxel can be approximated as a parallel plate capacitor described by~\cite{hayt_hayt1981engineering_1981}

\begin{equation}\label{capacitor}
    C_a = \displaystyle \frac{\epsilon_r \epsilon_0 A} {d} \; .
\end{equation}
where $C_a$ is the analog capacitance, $\epsilon_0$ is the vacuum permittivity, $\epsilon_r$ is the relative permittivity of the dielectric that covers the PCB, $A$ is the area of the taxels (equal for each taxel) and $d$ is the distance between the ground place and the taxels (Figure \ref{fig:scheme_insole}). 

The mechanical behaviour of the skin can be approximated as a set of parallel springs connecting the ground plane and the sensors plane. The behaviour can be described with Hooke's law~\cite{beatty_principles_1986} as:
\begin{equation}\label{Hook}
    F = - k (d - d_0) \; ,
\end{equation}
where $k$ is the stiffness of the skin and $d_0$ is the distance between the skin and the taxel, at rest conditions. The stiffness is approximated to a constant, thus, the model \eqref{Hook} is linear. The insole has a smooth, approximately flat shape. We assume the ground contact surface to be smooth as well, such that the insole deformation is locally flat around each taxel. Consequently, we consider only the  force perpendicular to the sensors. 

The force can also be expressed in terms of differential pressure relative to the atmospheric pressure, such that when no external force is applied to the skin (lifted foot) the differential pressure is zero. By assuming a uniform pressure $P$ over the taxel area $A$, the resulting force $F$ is given by:

\begin{equation}\label{force_pressure_over_area}
    F = A P \; .
\end{equation}
By substituting \eqref{Hook} and \eqref{force_pressure_over_area} in \eqref{capacitor}, the capacitance measured by each taxel is given by:

\begin{equation} \label{Ca}
    C_a = \displaystyle \frac{\epsilon_r \epsilon_0 A}{-\frac{PA}{k} + d_0} \; .
\end{equation}
By substituting \eqref{capacitor} and \eqref{Hook} in \eqref{force_pressure_over_area} and solving for $P$, we write the pressure estimated for each taxel as:

\begin{equation}\label{pressureFromCapacitance}
P = \frac{F}{A} = - \frac{k \left( d - d_0 \right)}{A} = - k \epsilon_r \epsilon_0 \left( \frac{1}{C_a} - \frac{1}{C_{a,0}} \right) \; .
\end{equation}
In \eqref{pressureFromCapacitance}, the pressure is linear with respect to $C_a^{-1}$ but the CDC circuits measure the analog capacitance $C_a$, convert and stream the capacitance as a raw value that we designate as digital capacitance.
The relationship between $C_a$ and $C_d$ is expressed by a linear function:

\begin{equation} \label{mappingDigital}
	C_d = \displaystyle \frac{C_{d,0}}{C_{a,0} - C_{a,max}} (C_a - C_{a,max}) \; .
\end{equation}
By considering the variation of both capacitances $C_a$ and $C_d$, the linear relationship is still valid, as follows:

\begin{equation} \label{CaCd}
	\Delta C_d = \displaystyle \frac{\Delta C_{d,max}}{\Delta C_{a,max}} \Delta C_a \; ,
\end{equation}
where
\begin{subequations}
	\begin{eqnarray}
		\Delta C_d &=& C_{d,0} - C_d  \label{DeltaCd} \; , \\
		\Delta C_a &=& C_{a} - C_{a,0} =  C_{a} - \displaystyle \frac{\epsilon_r \epsilon_0 A}{ d_0} \label{DeltaCa} \; .
	\end{eqnarray}
\end{subequations}
The variables $\Delta C_{d,max}$ and  $\Delta C_{a,max}$ are the maximum variation of the digital and analog capacitance, respectively.
%, $\Delta C_{d,max}$ is the maximum variation of the digital capacitance, $\Delta C_{a,max}$ is the maximum variation of the analog capacitance and $\Delta C_a$ is the measured variation of the analog capacitance and is given by $\Delta C_a = C_{a} - C_{a,0}$.
%
The substitution of \eqref{Ca}, \eqref{DeltaCd} and \eqref{DeltaCa} in \eqref{CaCd} yields the pressure exerted on each taxel as:

\begin{equation}\label{pressure}
P  =  \displaystyle   \frac   {d_{0} k} {A}  -   \displaystyle  \frac  { \epsilon_{r} \epsilon_0 k }   {   \displaystyle  \frac  { \Delta C_{a,max}}  { \Delta C_{d,max} } (C_{d,0} - C_d)  +  \displaystyle   \frac  {A \epsilon_{r} \epsilon_0  }  { d_{0} }   } \; .
\end{equation}
Nevertheless, the model in \eqref{pressure} can be applied only in ideal conditions. The variables $d_0$, $k$ and $\epsilon_{r}$ are typically obtained via a calibration procedure which is affected by several limitations:

\begin{itemize}
\item[-] $\epsilon_{r}$ changes with the pressure because the quantity of air contained between the ground plane and the sensors decreases to zero (we can consider that beyond this point we are saturating the sensors);
\item[-] $k$ changes with the pressure and with time;
\item[-] $d_0$ is not constant due to the hysteresis in the dielectric mechanical elasticity.
\end{itemize}
To overcome some of these drawbacks, a polynomial model can be used to describe the relationship between capacitance and pressure, thus overcoming some of the limitations of the model \eqref{pressure}. Hysteresis phenomena, however, will not be taken into account.

%%%%%%%%%%%%%%%%%%%%%%
\subsection{Recall on Skin Calibration} \label{subsec:calibration_joan}
The objective of this section is to present two different methods used in previous work for calibrating the iCub skin \cite{cannata_embedded_2008}. 
Both methods choose a fifth-order polynomial model and use setups based on a similar approach. They exploit an air pump and flexible material to uniformly distribute pressure on the surface of the skin. 

\begin{figure}[H]
	\centering
	\begin{minipage}{0.4\linewidth}
		\begin{subfigure}{\linewidth}\centering
			\includegraphics[width=0.976\linewidth]{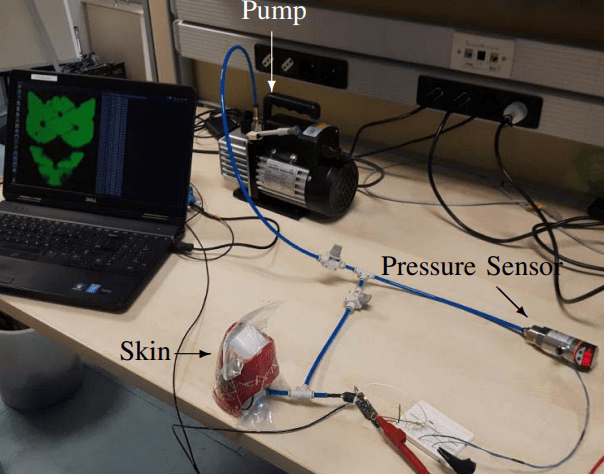}
			\caption{}
			\label{vacuum_bag_joan}
		\end{subfigure}
	\end{minipage}
	\begin{minipage}{0.55\linewidth}
		\begin{subfigure}{\linewidth}\centering
			\includegraphics[width=1\linewidth]{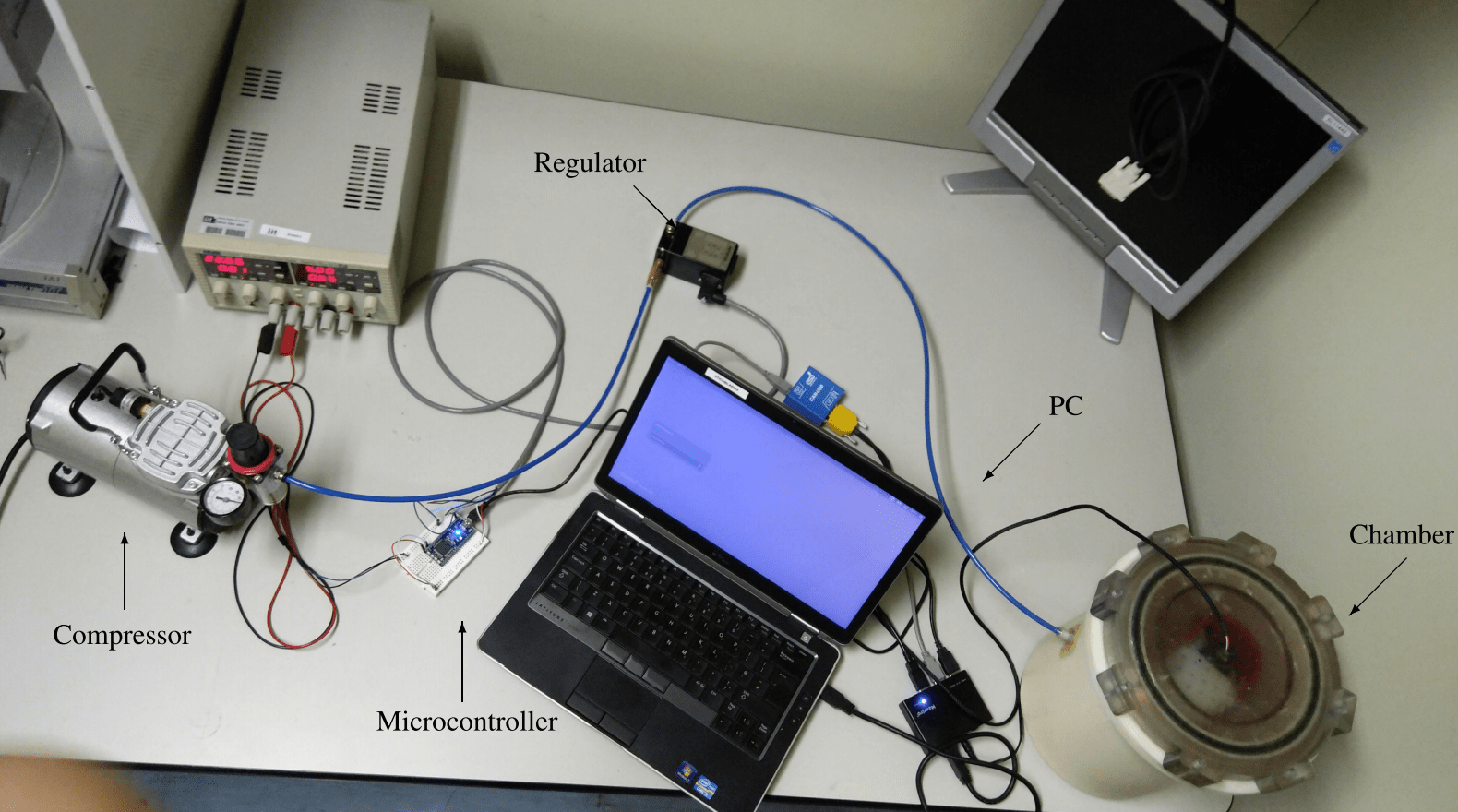}
			\caption{}
			\label{chamber_joan}
		\end{subfigure}
	\end{minipage}
	\caption{({\bf{a}}) Calibration setup based on vacuum bags. ({\bf{b}}) Calibration setup based on isolation chamber.}
\end{figure}

\begin{itemize}
	\item The first setup (Figure \ref{vacuum_bag_joan}) consists of inserting the skin inside a Ziploc bag and pumping out air. This decreases the pressure in the bag and creates a uniform pressure distribution over the skin surface. The flow rate is regulated by using valves to ensure a slow and steady pace in order to avoid the dynamic effects of the dielectric material. The pressure reached on the skin is measured with a pressure sensor that sends data to a computer through a CAN network~\cite{kangro_skin_2017}.
	\item The second setup uses an isolation chamber (Figure \ref{chamber_joan}). Contrary to the vacuum bag experiment explained above, the pressure is now increased in the external environment relative to the skin. The setup consists of an air compressor, the mentioned chamber, a pressure regulator and a micro-controller for pushing air and increasing the pressure inside the chamber~\cite{kangro_plenum-based_2018}.
\end{itemize}

In both cases, the capacitances measured by each taxel are related to the applied pressure by using a least-square fitting to find the polynomial coefficients.

In this paper, we use the first setup to take measurements then fused with data coming from sensorised shoes (Section \ref{subsec:theoretical_model_taxel}).

%% file: sections/contribution.tex
%!TEX root = ../IS_sensors2019.tex

\section{Contribution}\label{contribution}

This section describes our insole prototype by detailing the changes to the capacity-based technology recalled in Section \ref{subsec:hardware_skin}. It also presents a new setup for the insole calibration as well as the mathematical theory developed for the parameter identification for the taxels’ polynomial model. 

% ******************************************************************************

\subsection{Insole Prototype}\label{skin_insole}

\begin{figure}[H]
    \centering
    \begin{minipage}{0.39\linewidth}
        \begin{subfigure}{\linewidth}\centering
            \includegraphics[width=0.95\linewidth]{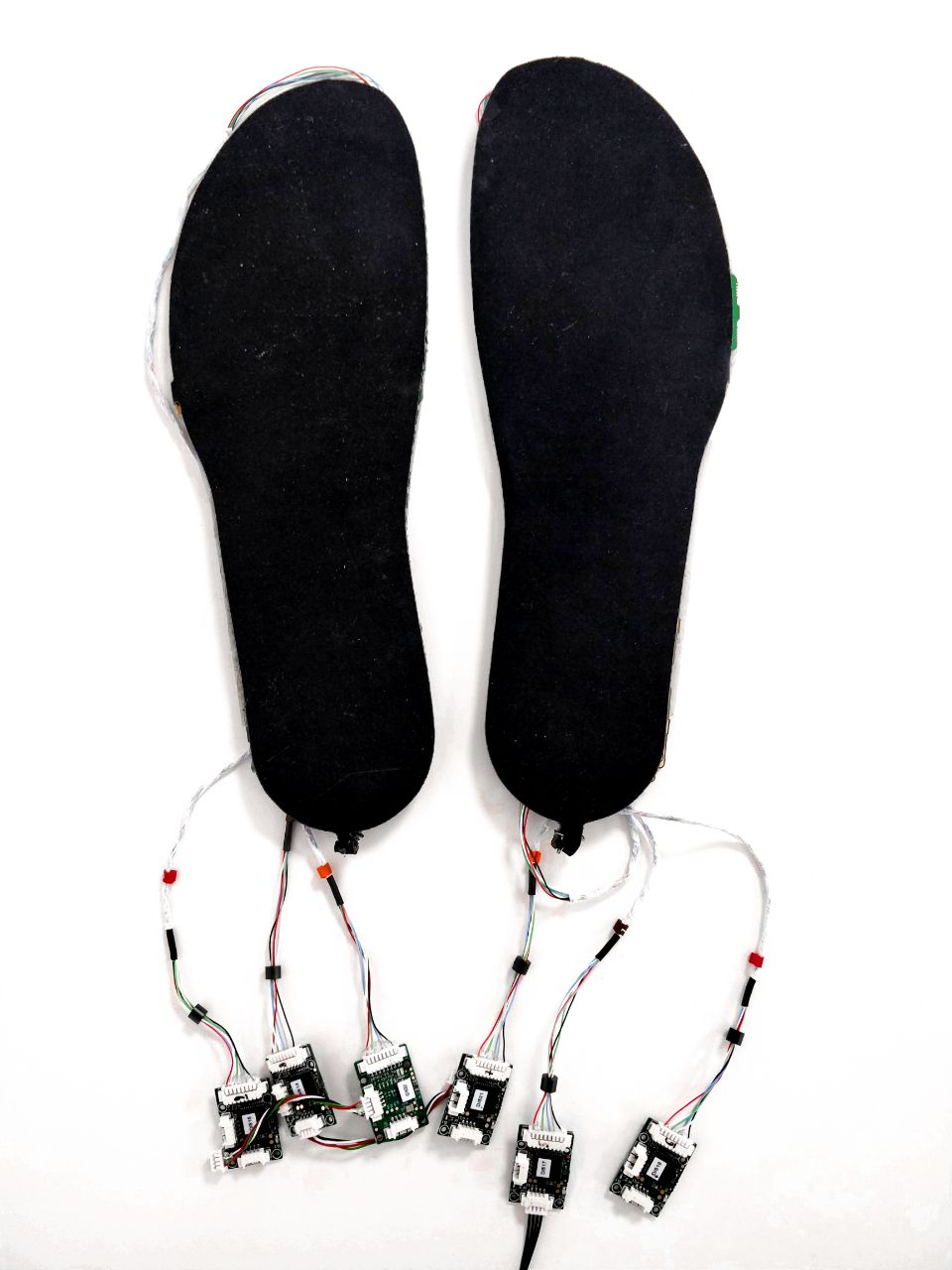}
            \caption{}
            \label{insoles_complete}
        \end{subfigure}
    \end{minipage}
    \begin{minipage}{0.2\linewidth}
        \begin{subfigure}{\linewidth}\centering
            \includegraphics[width=1\linewidth]{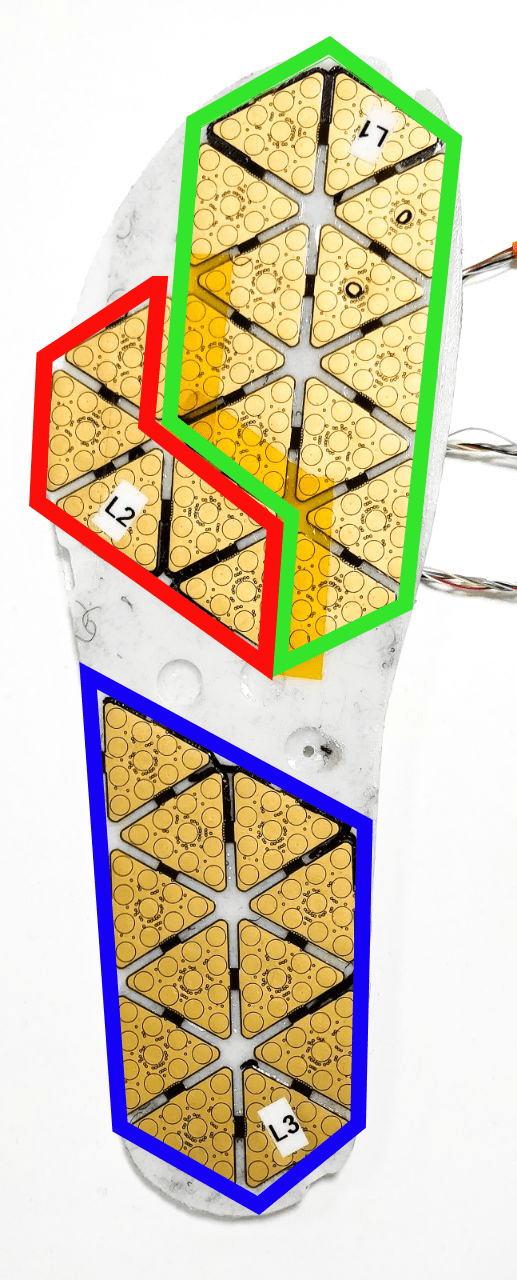}
            \caption{}
            \label{Insole_triangles}
        \end{subfigure}
    \end{minipage}
    \begin{minipage}{0.35\linewidth}
        \begin{subfigure}{\linewidth}\centering
            \includegraphics[width=0.98\linewidth]{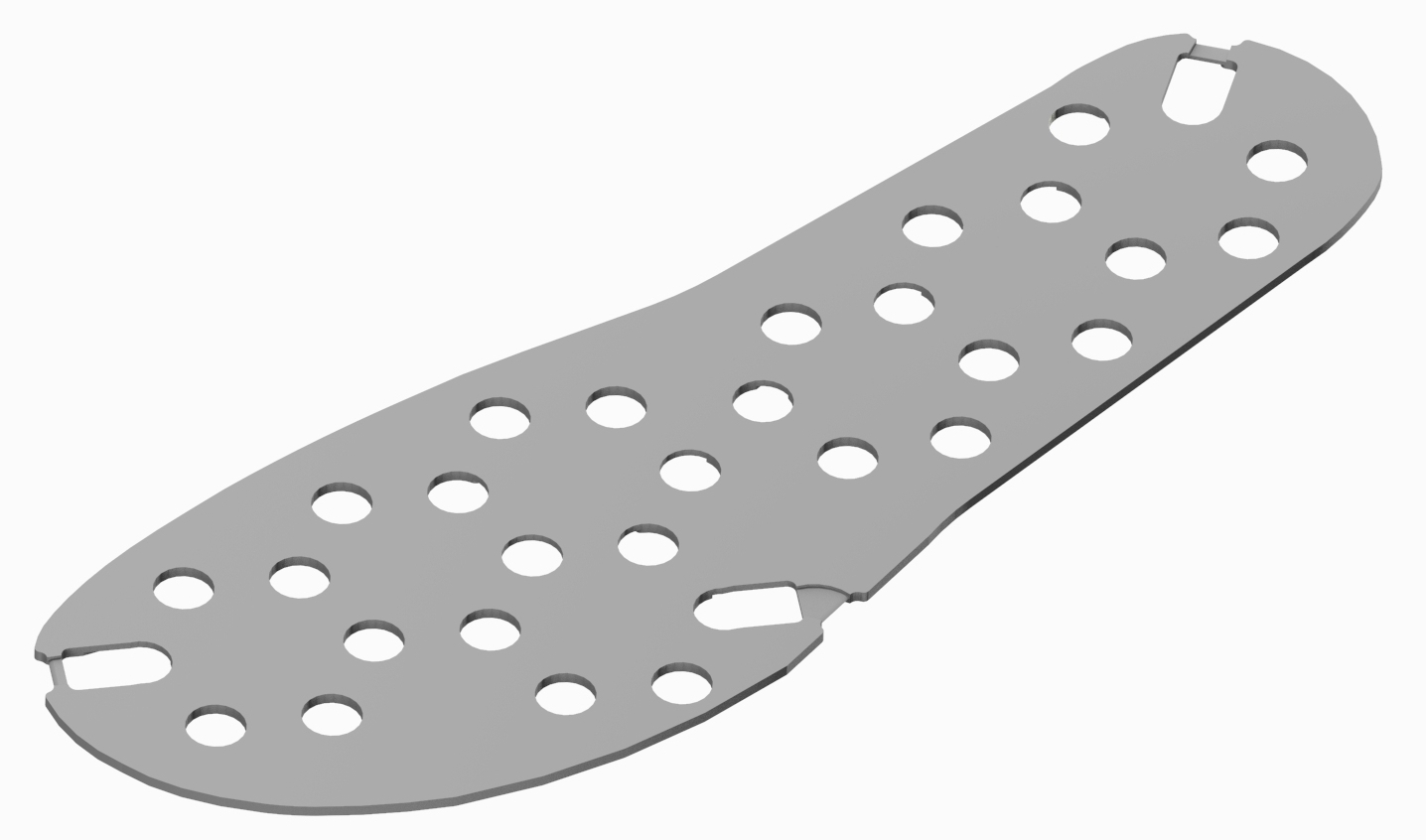}
            \caption{}
            \label{Insole_CAD}
        \end{subfigure}
        \vfill \vspace{1cm}
        \begin{subfigure}{\textwidth}\centering
            \includegraphics[width=0.55\linewidth]{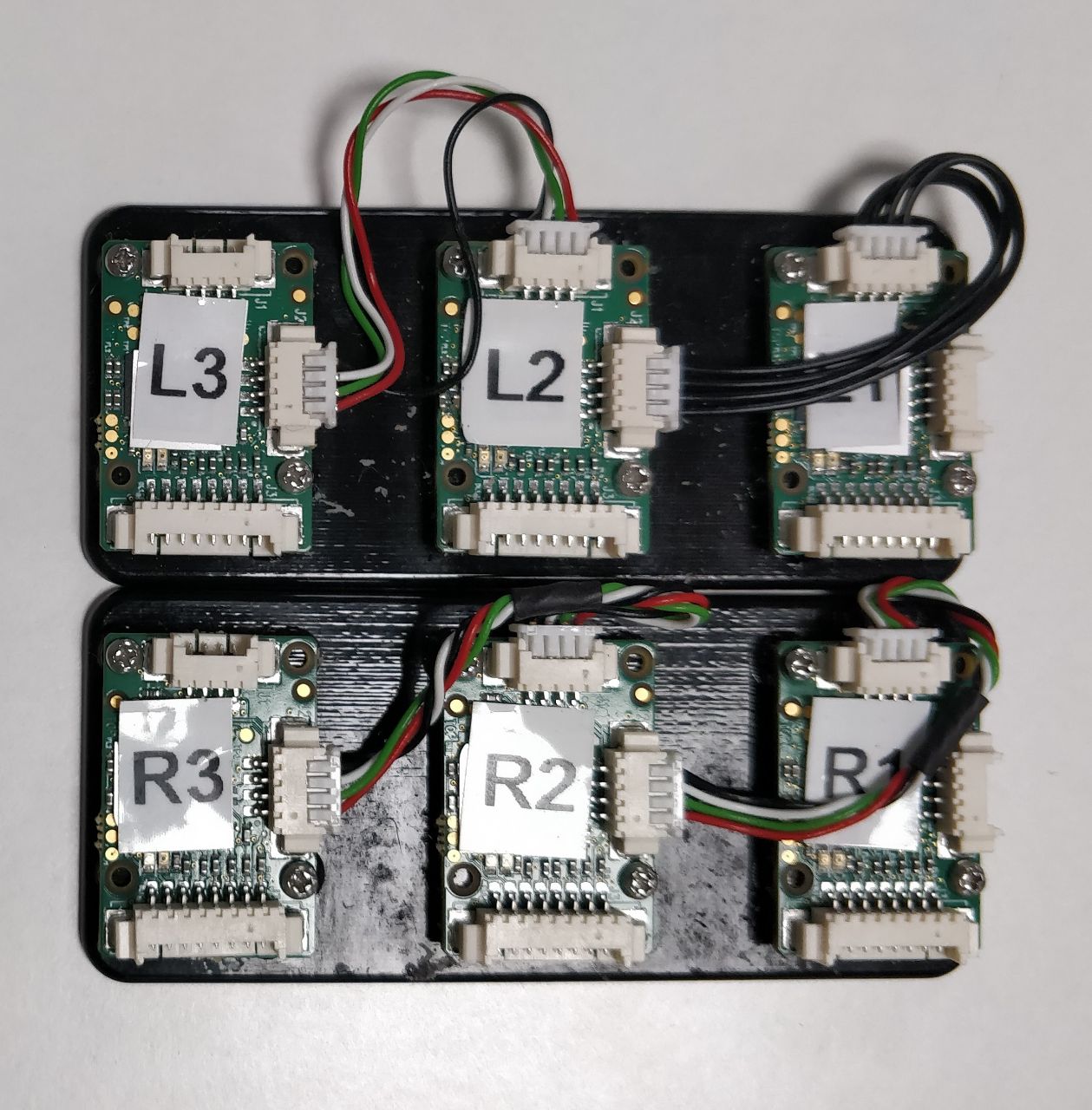}
            \caption{}
            \label{MTB}
        \end{subfigure}
    \end{minipage}
\caption{({\bf{a}}) Insoles prototype. ({\bf{b}}) Inner part of the insole containing the gaps that accommodate the digital converter integrated circuits. ({\bf{c}}) Computer-Aided Drafting (CAD)%Define if appropriate
~design of the support for the electronics. ({\bf{d}}) Micro-controller tactile boards (MTBs) consistency:  one for each patch of the insoles.}
\end{figure}

The insole prototype is based on the tactile capacitive technology presented in Section \ref{subsec:hardware_skin} (Figure \ref{insoles_complete}). The insole is mainly composed of four elements: a support, the electronics, a skin dielectric and the MTBs. The support is an element not present in the iCub skin, as the electronics are directly glued on the iCub cover. In our case, the electronic components are not directly glued on the internal base of the shoe, but on a dedicated support. This choice comes from the fact that the insole is continuously subjected to stress due to foot steps and high pressures, but at the same time the electronic components are very fragile and need to be protected. The support is a {3-mm-thick}%confirm 
~3D-printed piece of plastic% of 3 mm $3~\si{\milli\meter}$ thick 
~(Figure \ref{Insole_CAD}) with gaps to accommodate the CDC (Figure \ref{Insole_triangles}). The electronics of each insole are composed of 28 triangles, divided into three patches highlighted in Figure \ref{Insole_triangles}: front in green, middle in red and rear in blue. Each patch is connected to a different MTB. The MTBs are connected in series and they send information to a host computer connected to the insoles through a CAN bus interface. A total of 280 capacitive pressure sensors and 56 capacitive temperature sensors are available on each insole. The electronics are covered by the same cost-effective dielectric material used for the iCub skin, but made thicker to sustain the human weight range. The dielectric layer is covered by conductive lycra and is glued to the PCB (Figure \ref{insoles_complete}). The total weight of each insole is about $100~\si{\gram}$. The cost of a pair of insoles is not higher than 500 euros.

% ******************************************************************************
\subsection{New Calibration Method}\label{new_calibration_experiments}

% ***************** New calibration - Experiments *********************
As done in previous work~\cite{contact_ft_skin}, we are interested in estimating the pressure distribution, contact forces and moments using only our insole prototype. The relationship between these quantities and the digital capacitance is obtained through calibration. Nevertheless, the calibration methods described in Section \ref{subsec:calibration_joan} cannot be used for the calibration of our insole prototype owing to the following drawbacks:

%have drawbacks and cannot be used for the calibration of our insole prototype.

\begin{itemize}
	\item The vacuum bag method can reach a maximum pressure of 1 Atm, which is not high enough for our application. We need higher pressures during the calibration in order to take data that is consistent with real working conditions. For instance, if we consider a person weighing $80~\si{\kilogram\second}$%Confirm the units are 
~standing still on a single foot heel, a vertical force $f_z$ of about 800 N is applied on a small area of the insole. This force is spread over 4\textasciitilde 5 triangles. The area of each triangle is about $A = 3.24\times%Confirm
10^{-4}$ m$^2$. Thus, recalling \eqref{force_pressure_over_area} and considering four triangles, the pressure on each sensor will be:
	\begin{equation}
		P = \frac{f_z}{A*%Should asterisks used for multiplication be changed to \times throughout?
n_{triangles}} \approx \SI{6}{Atm} \; .
	\end{equation}
	\item The calibration setup based on the isolating vessel is not big enough to hold the proposed insole and can only reach a maximum pressure of 3 Atm.
\end{itemize}

Because of these drawbacks, we propose a new calibration setup and method. We combine data of two different setups for the insole calibration. The first setup is the vacuum bag. Although it does not allow pressures higher than 1 Atm to be reached, it allows to: $i)$ have a uniform pressure distribution over the insole; $ii)$ control the flow rate for increasing the pressure; $iii)$ measure the applied pressure as described in Section \ref{subsec:calibration_joan}. These characteristics are important for avoiding the dynamic effects of the dielectric and thus characterising only the static behaviour of the skin. The second setup is meant to overcome the limits of the vacuum bag method. The idea is to compare reliable and accurate force/moment measurements from third-party sensors with the respective quantities estimated through the insoles. Among those sensors there are sensorised treadmills, force plates and force/torque sensors. By placing the insole on one of the mentioned sensors, they can be used simultaneously and data from the insole and from force/moments sensors can be acquired in a synchronised way. This kind of setup allows higher pressures to be reached (depending on the weight of the person) and allows the taxels to be excited with different pressure distributions. The quantities that will be compared are: pressure, vertical ground reaction force and moments about horizontal axes.

% ***************** New calibration - Model approximation *********************
%% Mathematical formulation for insole calibration
%\subsection{Parameters Identification of Mathematical Model for Insole Calibration}
\subsection{Model Parameters Identification from the New Calibration Method}

The theoretical model of taxels described in Section \ref{subsec:theoretical_model_taxel} considers ideal behaviours and precise knowledge of quantities like the skin stiffness $k$ and the dielectric constant $\epsilon_r$. In practice, these values are difficult to estimate. 
We resort to a polynomial function for describing taxels' calibration curves for two reasons:
\begin{itemize}
    \item[-] The quantities $d_0$ and $\epsilon_r$ to estimate in the model \eqref{pressure} are not linear with respect to the pressure and to find them we need to solve a non-linear optimisation problem;
    \item[-] Hypotheses for deriving the theoretical model are based on ideal behaviours, but the physical insole does not always respect the ideally hypotised case. Thus, a more general model helps to consider factors that we did not take into account in \eqref{pressure}.
\end{itemize}

We use a third-degree polynomial, where the degree was chosen on the basis of the sensors' answer. We checked the shape of the taxel calibration curves relating the pressure $P(C_i)$ and digital capacitances $C_i$ acquired by increasing the pressure with the vacuum bag setup. Indeed, the calibration curves  often have two inflexions due to a nonlinearity at low differential pressure and near saturation of the sensor, and the two inflexions are typical of a cubic function. Thus, the model for the $i$-th taxel is defined as:

\begin{equation}
    P(C_i) = a_{0,i} + a_{1,i} C_i + a_{2,i} C_i^2 + a_{3,i} C_i^3 \; .
\end{equation}
The model may be different for each taxel, and the individual coefficients are estimated through the calibration procedure. It is possible to write each taxel model in a compact form:

\begin{equation}
    P(C_i) = \boldsymbol{\gamma}_i^T \boldsymbol{k}_i \; ,
\end{equation}
where $\boldsymbol{\gamma}_i = \begin{bmatrix} 1 & C_i & C_i^2 & C_i^3 \end{bmatrix}^T$ contains the powers of the capacitance $C_i$ for the sensor $i$, and $\boldsymbol{k}_i = \begin{bmatrix} a_{0,i}  & a_{1,i} & a_{2,i} & a_{3,i} \end{bmatrix}^T$ contains the respective polynomial coefficients to estimate.

The resultant force acting on the insole is computed considering the force on each taxel:

\begin{equation}\label{force_z_sk}
    \boldsymbol { f }^{SK} = \displaystyle \sum_i {P_i A_i} \boldsymbol {\hat n} \; ,
\end{equation}
where $\boldsymbol { f }^{SK} \in \mathbb{R}^{3}$, and for each taxel $i$, $A_i$ is the taxel area on the ground and $P_i$ is the measured pressure. $\boldsymbol {\hat n}$ is the unit vector normal to the ground and pointing upwards, since each taxel measures a force normal to the sensor plate and we consider both the insole and the ground contact to be flat. 
Another direct consequence is that it is not possible to compute the ground contact horizontal forces. In addition, all the taxels have the same known area denoted by $A$.
For compactness in the formulation, we consider a reference frame $W$ with its origin placed anywhere on the ground plane, the $x$ and $y$ axes parallel to the ground plane, the $z$ axis pointing up. All the dynamic quantities and taxel positions are expressed in that frame. We have $\boldsymbol {\hat n} = \begin{bmatrix}0 & 0 & 1\end{bmatrix}^T$ such that we can consider just the \textit{z} component of the normal measured forces:
\begin{equation}
    f_z^{SK} = \displaystyle \sum_i {P_i A_i} = A \displaystyle \sum_i { \boldsymbol{\gamma}_i^T \boldsymbol{k}_i} \; ,
\end{equation}
being $f_z^{SK} \in \mathbb{R}$. By defining $\boldsymbol{\varphi}_{f} \in \mathbb{R}^{4n \times 1}$ as:

\begin{equation}
    \boldsymbol{\varphi}_{f} = \begin{bmatrix} \boldsymbol{\gamma}_1^T & \boldsymbol{\gamma}_2^T & \cdots & \boldsymbol{\gamma}_n^T \end{bmatrix}^T \; ,
\end{equation}
where $n$ is the number of taxels that compose the insole, and $\boldsymbol{k} \in \mathbb{R}^{4n \times 1}$ as:

\begin{equation}
    \boldsymbol{k} = \begin{bmatrix} \boldsymbol{k}_1^T & \boldsymbol{k}_2^T & \cdots & \boldsymbol{k}_n^T \end{bmatrix}^T \; ,
\end{equation}
we can write the resultant force as:

\begin{equation}
    f_z^{SK} = A \boldsymbol{\varphi}_{f}^T \boldsymbol{k} \; .
\end{equation}
Let $\boldsymbol{X} \in \mathbb{R}^{4n \times 4n} $ and $\boldsymbol{Y} \in \mathbb{R}^{4n \times 4n} $ denote the block diagonal matrices containing respectively the \textit{x} and \textit{y} coordinates of each taxel expressed in the reference frame $W$, in the form:

\begin{equation}
    \boldsymbol{X} = 
    \begin{bmatrix}  
        x_1 I_4  &  &  \\
        &  \ddots  &  \\
        &  &  x_n I_4
    \end{bmatrix},
    \quad
    \boldsymbol{Y} = 
    \begin{bmatrix}  
        y_1 I_4  &  &  \\
        &  \ddots  &  \\
        &  &  y_n I_4
    \end{bmatrix} \; .
\end{equation}
We can then write the resultant moments as:

\begin{equation}\label{moments_sk}
    m_x^{SK} = A \boldsymbol{\varphi}_{f}^T \boldsymbol{Y} \boldsymbol{k} \; ,
    \quad
    m_y^{SK} = A \boldsymbol{\varphi}_{f}^T \boldsymbol{X} \boldsymbol{k} \; .
\end{equation}
The number of samples acquired is greater than the number of unknown coefficients. Thus, it is possible to define four over-constrained systems in the following form:

\begin{equation}\label{LS}
\boldsymbol \Phi \boldsymbol k = \boldsymbol b \; ,
\end{equation}
where $\boldsymbol \Phi \in \mathbb{R}^{m \times 4n}$ is the regressor containing the capacitances' powers measured by the taxels, $\boldsymbol k \in \mathbb{R}^{4n}$ are the coefficients to estimate and $\boldsymbol b \in \mathbb{R}^{m}$, with $m \gg 4n$, containing the known quantities measured by the accurate sensors. In general, a problem in the form $\boldsymbol \Phi \boldsymbol k = \boldsymbol b$ can be solved through the least-square method.
Nevertheless, in our case there are multiple variables to be compared, that is, vertical ground reaction force, moments about horizontal axes and pressure. 

\noindent The optimisation problem could be ill-posed for two reasons:
\begin{itemize}
    \item[-] Variables to compare, force, moments and pressure have different order. Indeed, force can reach a value of about $1000~\si{\newton}$, moments can reach a value of about $100~\si{\newton\meter}$ and pressure assumes values between 0 and $10^6~\si{\pascal}$.
    \item[-] The sample numbers of the two data sets (FT shoes data set and vacuum bag data set) is not the same because they come from different experiments.
\end{itemize}

Thus, we need to normalise the variables of the optimisation problem (forces, moments and pressures) by considering the order of magnitude and the sample numbers of each data set. The normalisation is obtained by adopting the weighted least-square method:
\begin{equation}
\begin{split}
   \boldsymbol k^* &= \operatorname*{argmin}_{\boldsymbol k} \Big ( w_{f} || \boldsymbol \Phi_{f} \boldsymbol k - \boldsymbol{f}_z^{FT} || ^2 + w_{m_x} || \boldsymbol \Phi_{m_x} \boldsymbol k - \boldsymbol{m}_x^{FT} || ^2 \\
&+ w_{m_y} || \boldsymbol \Phi_{m_y} \boldsymbol k - \boldsymbol{ m}_y^{FT} || ^2  + \sum_{i} { w_{P,i} ||  \boldsymbol \Phi_{P,i} \boldsymbol k - \boldsymbol P || ^2 }  \Big )  \; ,
\end{split}
\end{equation}
where $\boldsymbol{f}_z^{FT} \in \mathbb{R}^{l_{f}}$, $\boldsymbol{m}_x^{FT}  \in \mathbb{R}^{l_{f}}$, $\boldsymbol{m}_y^{FT}  \in \mathbb{R}^{l_{f}}$ are measured by the FT sensors, $\boldsymbol P  \in \mathbb{R}^{l_P}$ is measured by the pressure sensor and they correspond to $\boldsymbol{b}$ in \eqref{LS}. $l_{f}$ and $l_P$ are the respective numbers of samples. Instead, regressors are defined as below:

\begin{subequations}
	\begin{eqnarray}
    \boldsymbol \Phi_{f} &=& \begin{bmatrix} A \boldsymbol{\varphi}_{f,1} & A \boldsymbol{\varphi}_{f,2} & \cdots & A \boldsymbol{\varphi}_{f,l_{f}} \end{bmatrix}^T \in \mathbb{R}^{l_{f} \times 4n} \; , \\
    \boldsymbol \Phi_{m_x} &=& \begin{bmatrix} A \boldsymbol{\varphi}_{f,1} \boldsymbol{X} & A \boldsymbol{\varphi}_{f,2} \boldsymbol{X} & \cdots & A \boldsymbol{\varphi}_{f,l_{f}} \boldsymbol{X} \end{bmatrix}^T \in \mathbb{R}^{l_{f} \times 4n} \; , \\
    \boldsymbol \Phi_{m_y} &=& \begin{bmatrix} A \boldsymbol{\varphi}_{f,1} \boldsymbol{Y} & A \boldsymbol{\varphi}_{f,2} \boldsymbol{Y} & \cdots & A \boldsymbol{\varphi}_{f,l_{f}} \boldsymbol{Y} \end{bmatrix}^T \in \mathbb{R}^{l_{f} \times 4n} \; , \\
    \boldsymbol \Phi_{P,i} &=& \begin{bmatrix} \boldsymbol{0}_{{m_{P},i} \times 4(i-1)} & \begin{matrix} \boldsymbol{ \gamma}_{i,1}^T \\ \vdots \\ \boldsymbol{\gamma}_{i,l_{P}}^T \end{matrix} & \boldsymbol{0}_{{m_{P},i} \times 4(n-i)}  \end{bmatrix} \in \mathbb{R}^{l_P \times 4n} \; .
	\end{eqnarray}
\end{subequations}
The problem can also be written in a quadratic programming form:

\begin{equation}
\boldsymbol k^* = \operatorname*{argmin}_{\boldsymbol k} \bigg ( \frac{1}{2} \boldsymbol{k}^T \boldsymbol{H} \boldsymbol{k} + \boldsymbol{k}^T \boldsymbol{g} \bigg ) \; ,
\end{equation}
where

\begin{subequations}
	\begin{eqnarray}
	\boldsymbol{H} &=& w_{f}  \boldsymbol \Phi_{f}^T  \boldsymbol \Phi_{f} + w_{m_x}  \boldsymbol \Phi_{m_x}^T  \boldsymbol \Phi_{m_x} + w_{m_y} \boldsymbol \Phi_{m_y}^T \boldsymbol \Phi_{m_y} +  \sum_{i} { w_{P,i} \boldsymbol \Phi_{P,i}^T  \boldsymbol \Phi_{P,i}  }  \; , \\
	\boldsymbol{g} &=& - w_{f}  \boldsymbol \Phi_{f}^T \boldsymbol{f}_z^{FT} - w_{m_x}  \boldsymbol \Phi_{m_x}^T  \boldsymbol{ m}_x^{FT} - w_{m_y} \boldsymbol \Phi_{m_y}^T \boldsymbol{ m}_y^{FT} - \sum_{i} { w_{P,i} \boldsymbol \Phi_{P,i}^T \boldsymbol{P} }  \; .
	\end{eqnarray}
\end{subequations}
The problem could be ill-posed, considering the nature of the problem and the fact that with a sensorised shoes setup we do not have control of the acquired data. To overcome this issue, the problem can be regularised with a Tikhonov regularisation~\cite{Neubauer_1989} as follows:

\begin{equation}\label{regulatization}
\begin{split}
\boldsymbol k^* &=  \operatorname*{argmin} \Big ( w_{f} || \boldsymbol \Phi_{f} \boldsymbol k - \boldsymbol{f}_z^{FT} || ^2 + w_{m_x} || \boldsymbol \Phi_{m_x} \boldsymbol k - \boldsymbol{ m}_x^{FT} || ^2 \\
&+w_{m_y} || \boldsymbol \Phi_{m_y} \boldsymbol k - \boldsymbol{ m}_y^{FT} || ^2  + \sum_{i=1}^{280} { w_{P_i} || \boldsymbol \Phi_{P_i} \boldsymbol k - \boldsymbol P || ^2 }  +  \boldsymbol \lambda  || \boldsymbol k ||^2 \Big ) \; .
\end{split} 
\end{equation}
The above quadratic programming problem is solved using an open-source C++ software package \textit{qpOASES}~\cite{Ferreau2014}. The main advantage of this tool is the possibility of adding constraints to our problem if required.

%% file: sections/experiments_and_validation.tex
%!TEX root = ../IS_sensors2019.tex

\section{Experiments and Validation}\label{experiments_and_validation}

%%%%%%%%%%%%%%%%%%%%%%
\vspace{-0.3cm}
\subsection{Experiments for Calibration and Validation}\label{experiments}
The objective of the experiments was $i)$ to collect data needed for the calibration of the taxels and $ii)$ to validate the effectiveness of the identification algorithm in \eqref{regulatization}. An experimental session was carried out at the Istituto Italiano di Tecnologia (IIT), Genova, Italy. One volunteer (mass: $63~\si{\kilo\gram}$) was recruited for the experiments to acquire the calibration dataset and a second volunteer (mass: $54~\si{\kilo\gram}$) was recruited for validating the estimation algorithm outcome.
%\pakosays{not entirely convinced about the goodness word, sounds a bit subjective while we are looking for something objective. Maybe validating the fitting of the estimation algorithm or the effectiveness of the calibration}

For collecting data to use at the calibration phase, two kinds of data sets were taken. 

\begin{figure}[H]
	\begin{subfigure}[t]{0.39\textwidth}\centering
		% include first image
		\includegraphics[width=0.9\columnwidth]{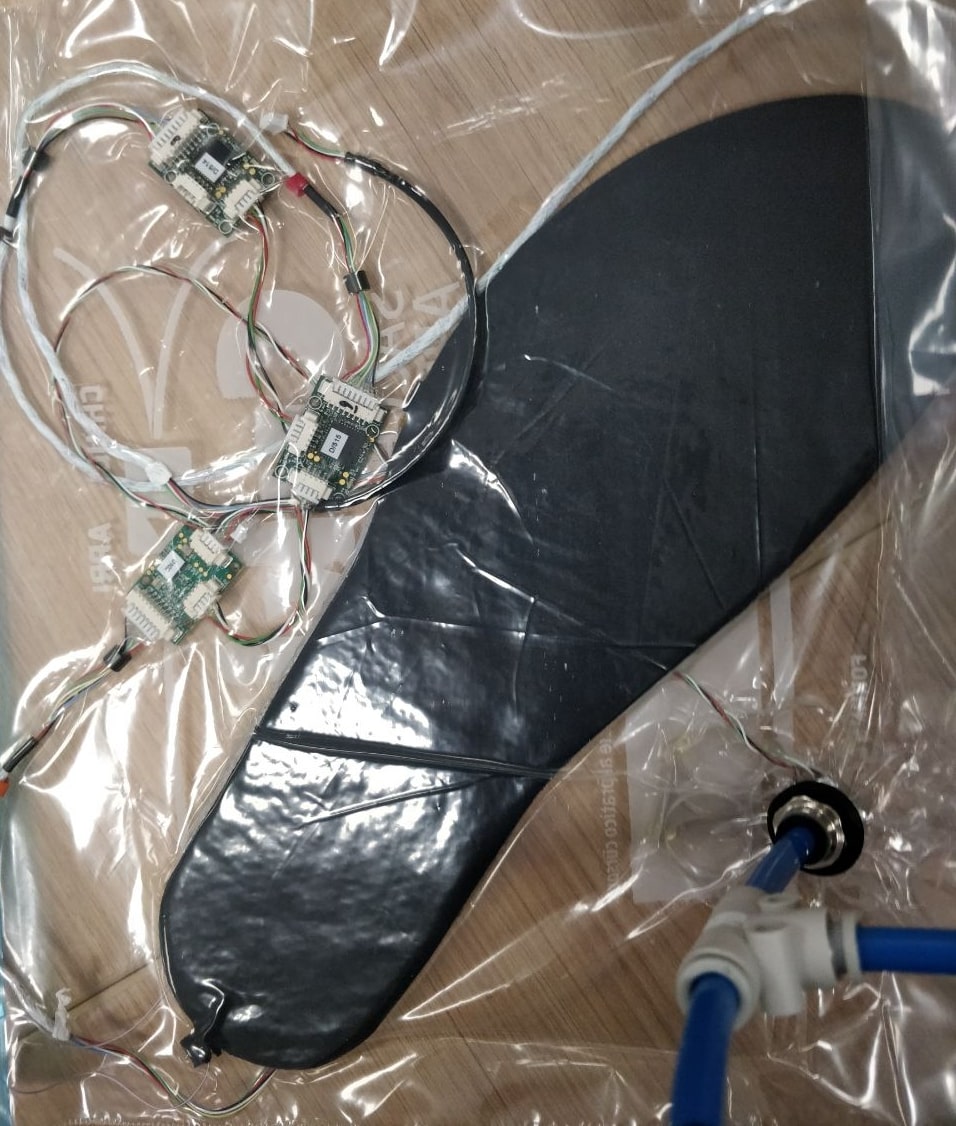}  
		\caption{}
		\label{fig:vacuum_bag}
	\end{subfigure}
	\begin{subfigure}[t]{0.59\textwidth}\centering
		% include second image
		\includegraphics[width=0.94\columnwidth]{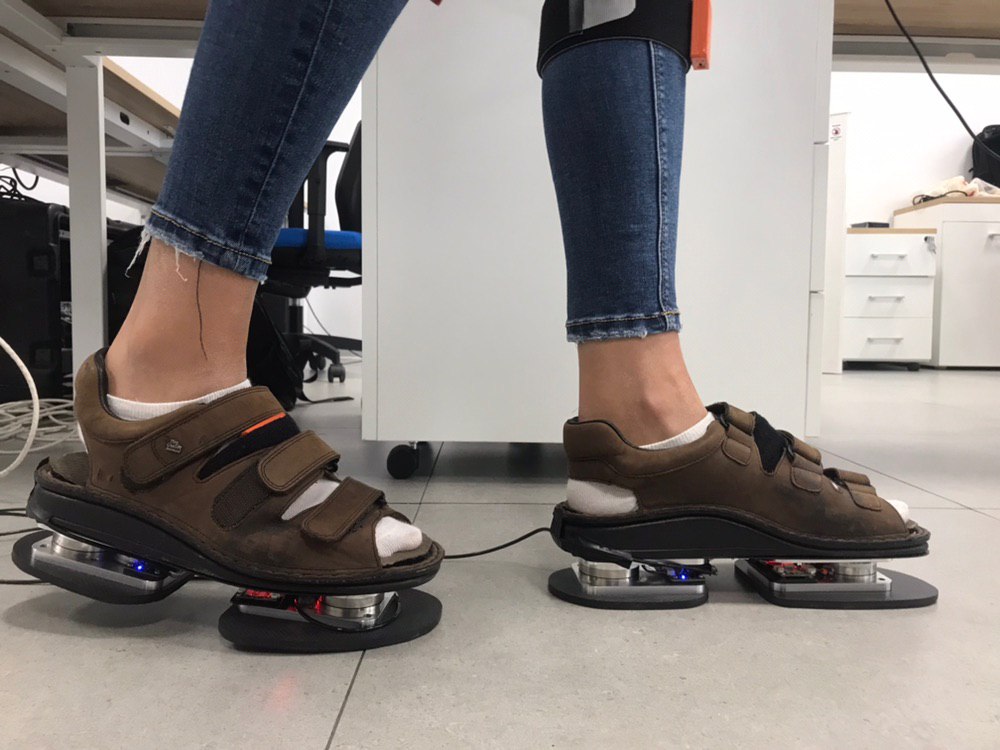}  
		\caption{}
		\label{fig:ftshoe}
	\end{subfigure}
	\caption{({\bf{a}}) Vacuum bag experiment. The pressure inside the bag was reduced by using a vacuum pump. ({\bf{b}}) Sensorised shoes equipped with force/torque (FT) sensors and insoles.}
	\label{fig:new_insole_and_tango_material}
\end{figure}

\begin{figure}[H]
\centering
	\includegraphics[width=\columnwidth]{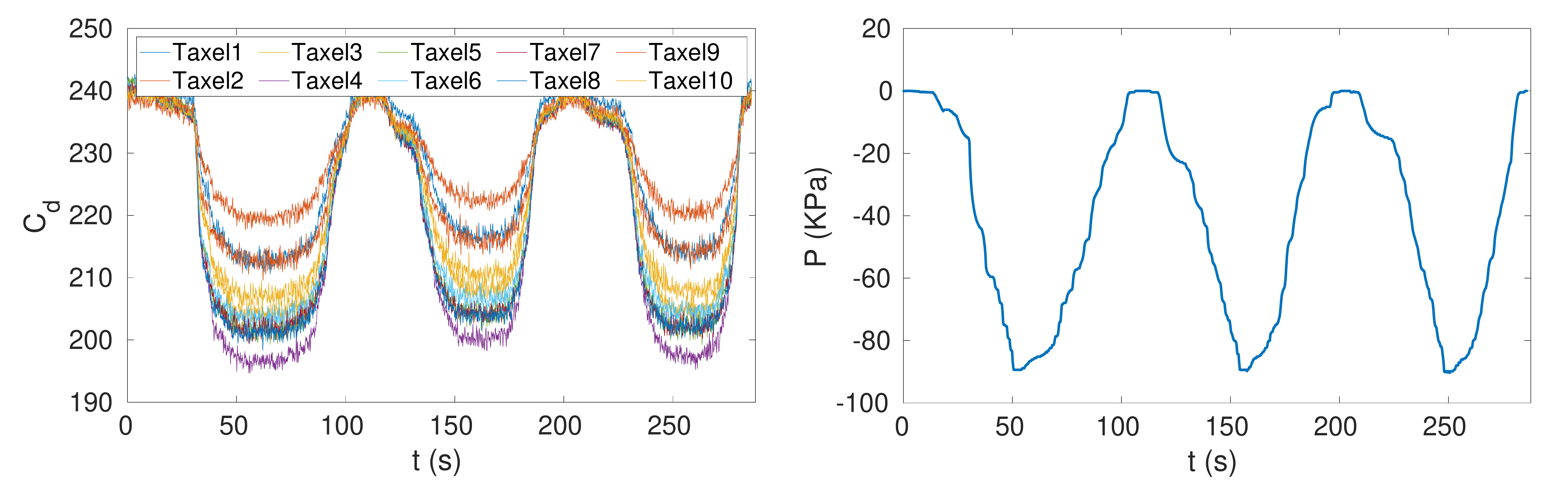}  
	\caption{Capacitances measured by one triangle (i.e., 10 taxels) and pressures measured by the pressure sensor during the vacuum bag experiment that consisted of three cycles of pressure reduction and increase.}%Correct "KPa" to kPa" on the axis
	\label{fig:vacuum_experiment}
\end{figure}

\begin{figure}[H]
\centering
	\includegraphics[width=\columnwidth]{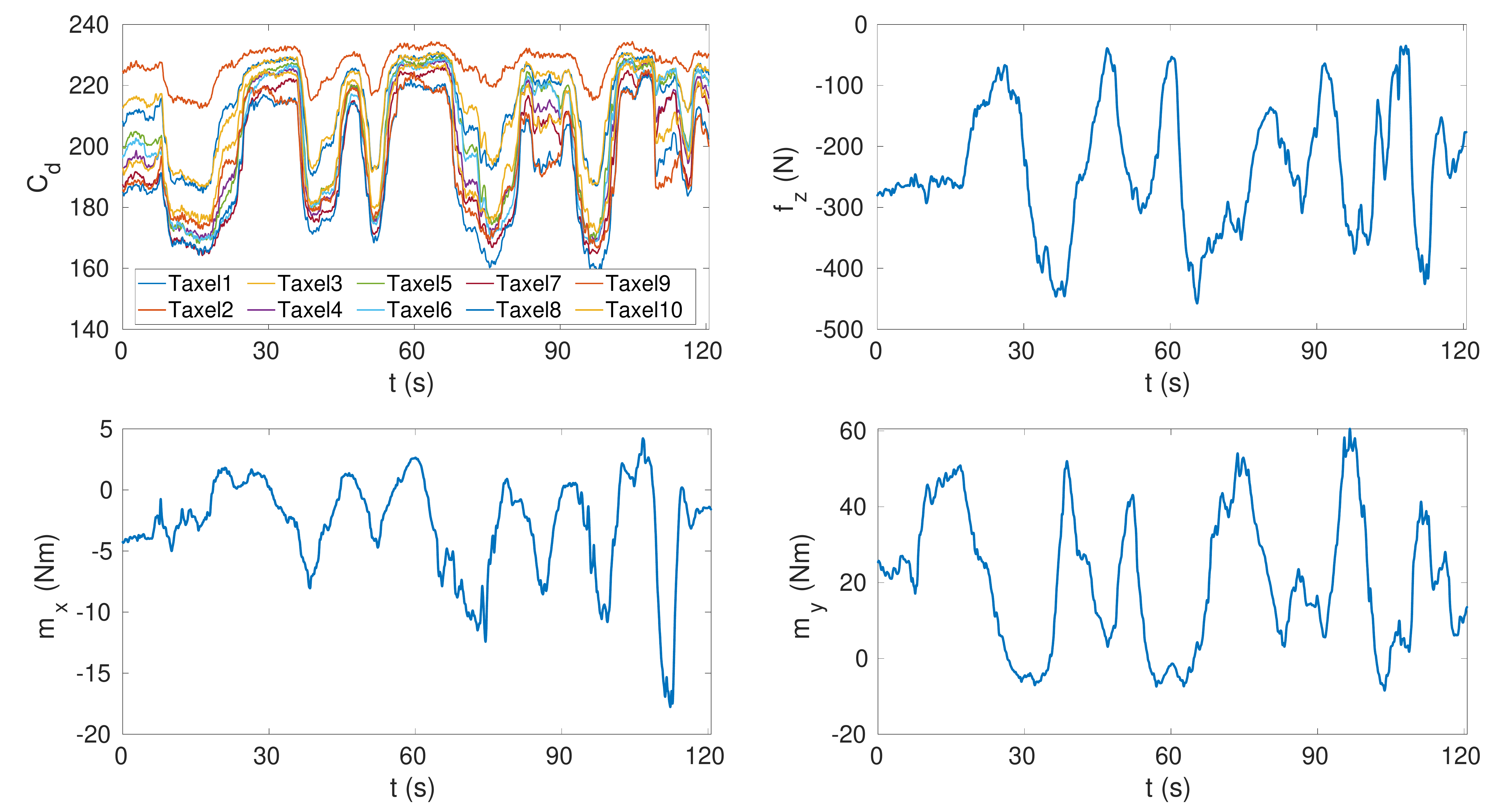}  
	\caption{Capacitances measured by one triangle (i.e., 10 taxels), vertical force and horizontal moments measured by the sensorised shoes.}
	\label{fig:ftshoesexp}
\end{figure}

\begin{itemize}
    \item One data set was collected using the setup described in Section \ref{subsec:calibration_joan}. The experiments with a vacuum bag (Figure \ref{fig:vacuum_bag}) reached a pressure of 90 kPa and the pressure was uniformly distributed over the insole. The experiment consisted of three cycles of pressure reduction and increase (Figure \ref{fig:vacuum_experiment}). In Figure \ref{fig:vacuum_experiment} we can see that the pressure was negative, which can be explained by the choice of the reference frame $W$ that has the $z$ axis pointing up. Indeed, in order to have compression, the force applied on each taxel has to be negative and as consequence of \eqref{force_pressure_over_area} even the pressure has to be negative. We can even observe that when the pressure increased the measured capacitance decreased. This is because the chips do not stream analog capacitances, as explained in Section \ref{subsec:theoretical_model_taxel}, but the analog capacitances are mapped to a digital value at an 8-bit resolution and the relation \eqref{mappingDigital}. The mapping was inverse, that is, values from 255 to 240 indicate that the taxel was not pressed and 0 means that the taxel was measuring the highest analog capacitance that is possible to map with 8 bits.

    \item Another data set was taken using a pair of sensorised shoes developed at IIT (Figure \ref{fig:ftshoe}). Each shoe was equipped with two six-axis FT sensors by the iCub group at Istituto Italiano di Tecnologia (IIT) \cite{ftsensors}, able to measure the forces and moments. As mentioned in Section \ref{new_calibration_experiments}, the data sets had different features in order to excite the activation of the taxels as much as possible in order to obtain different combinations to cover the entire capacitance range of each sensor (Figure \ref{fig:ftshoesexp}).
%    as more as possible the activation of the taxels to obtain different combination of activation and
\end{itemize}

%\iffalse
%\fi

%\iffalse
%\fi

%\iffalse
%\fi

\begin{figure}[H]
\centering
	\begin{subfigure}[t]{0.195\textwidth}\centering
		% include first image
		\includegraphics[width=0.9\columnwidth]{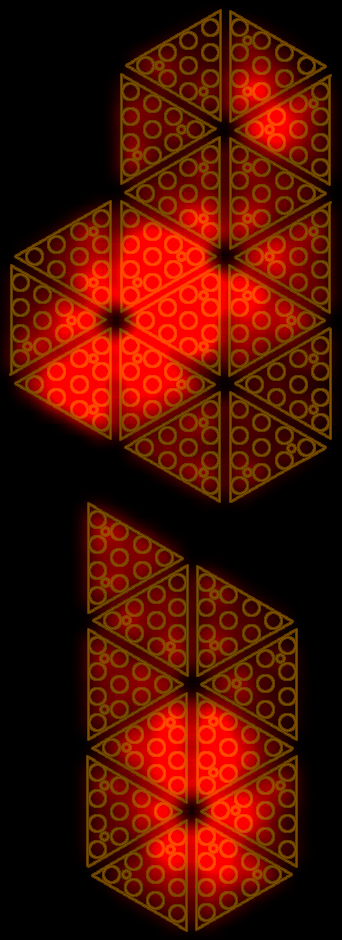}  
	\end{subfigure}
	\begin{subfigure}[t]{0.195\textwidth}\centering
		% include first image
		\includegraphics[width=0.9\columnwidth]{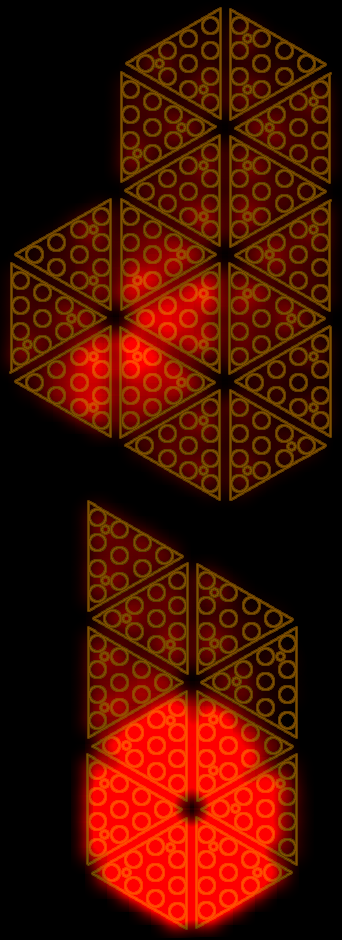}  
	\end{subfigure}
	\begin{subfigure}[t]{0.195\textwidth}\centering
		% include first image
		\includegraphics[width=0.9\columnwidth]{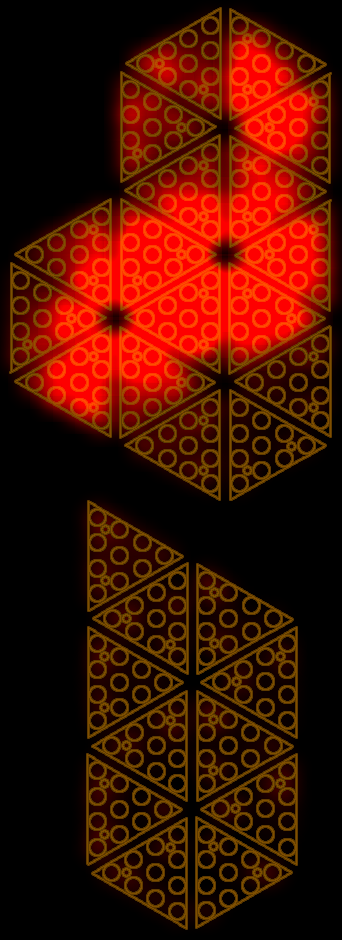}  
	\end{subfigure}
	\begin{subfigure}[t]{0.195\textwidth}\centering
		% include first image
		\includegraphics[width=0.9\columnwidth]{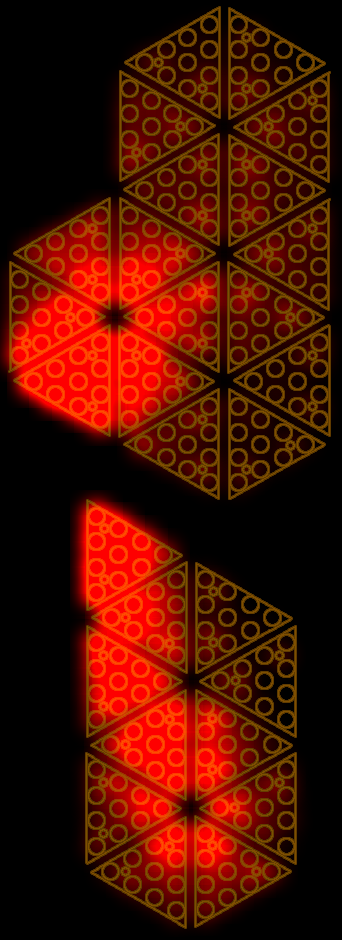}  
	\end{subfigure}
	\begin{subfigure}[t]{0.195\textwidth}\centering
		% include first image
		\includegraphics[width=0.9\columnwidth]{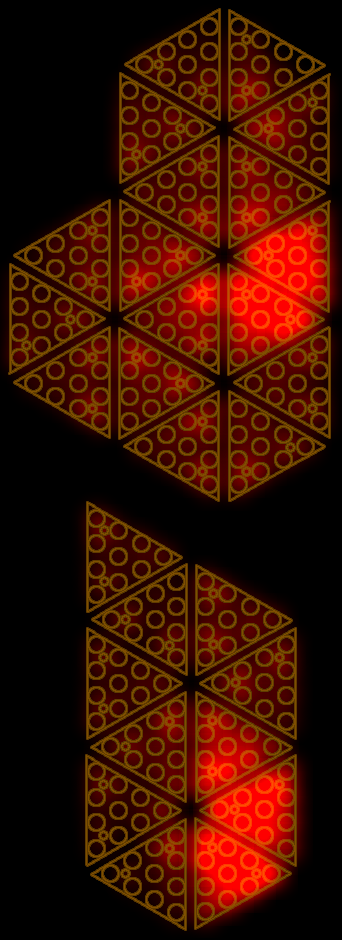}  
	\end{subfigure}
	\caption{ Taxel activation of the left insole in the experiment with sensorised shoes.}
	\label{fig:experiment_shoes}
\end{figure}

Several tasks were performed with the sensorised shoes, including standing still on both feet (T1), standing still on one foot (T2), performing slow movements on the feet (T3) (used for the calibration) and walking (T4--T6) (used for the validation). The pressure reached with these experiments was higher than the one applied on the insole using the vacuum bag (up to 1 MPa) and was not uniformly distributed (Figure \ref{fig:experiment_shoes}). The data sets used for calibration were pre-processed in order to filter the acquisition noise. The filter we used in pre-processing phase was the MATLAB implementation of the Savitzky--Golay filter \cite{sgolay}. The filter parameters were the order and the frame length and we set them to 5 and 41, respectively.
Data were processed and analysed with MATLAB\textsuperscript{\tiny\textregistered}%The registered trademark symbol is not given to every instance of MATLAB in this paper. Check that consistency and correctness is maintained throughout
. %\pakosays{maybe add a reference?}

The results described in the next paragraph were obtained from the left insole. The data sets for the calibration phase were filtered in order to consider only slowly varying data. The reason for this is that, compared to the insole sensors, the FT sensors and the pressure sensor are more responsive to rapid force/moment/pressure changes, which leads to a mismatch in the compared data and might affect the calibration results.

Let $n_f$ and $n_P$ be the number of samples in the data set collected with the sensorised shoe and the number of samples collected with the vacuum, respectively. The weights of the least-square problem were manually tuned and chosen as $w_{f} = 1/n_f$, $w_{m_x} = 33/n_f$, $w_{m_y} = 10/n_f$ and $w_{P_i} = 10^{-6}/n_P$, with $i=1,...,n$, where $n$ is the number of taxels on a single insole. We obtained the best performances with the chosen weights. The regularisation coefficient $\lambda$ was instead set to 10$^{-9}$. The highest force and moments reached were $|f_z| = 650~\si{\newton}$, $|m_x| = 25~\si{\newton\meter}$ and $|m_y| = 90~\si{\newton\meter}$.

%%%%%%%%%%%%%%%%%%%%%%
\subsection{Validation}\label{method}

\begin{figure}[H]
	\centering
	\begin{subfigure}[b]{1\textwidth}
		\includegraphics[width=1\linewidth]{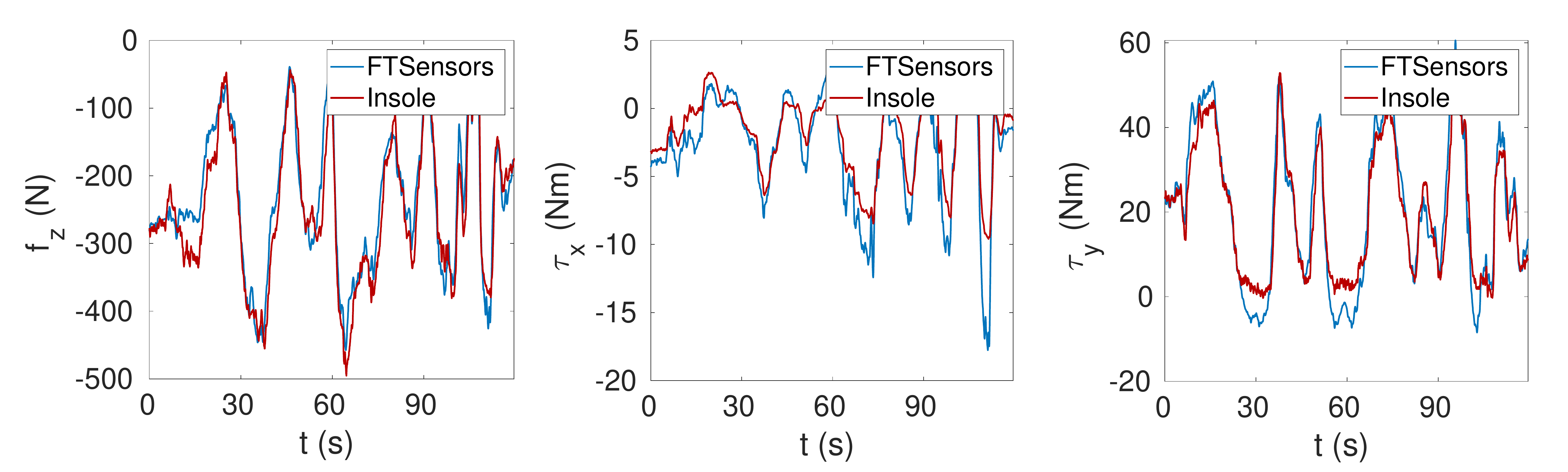}
		\caption{}
		\label{fig:slow_0_tracking} 
	\end{subfigure}
	
	\begin{subfigure}[b]{1\textwidth}
		\includegraphics[width=1\linewidth]{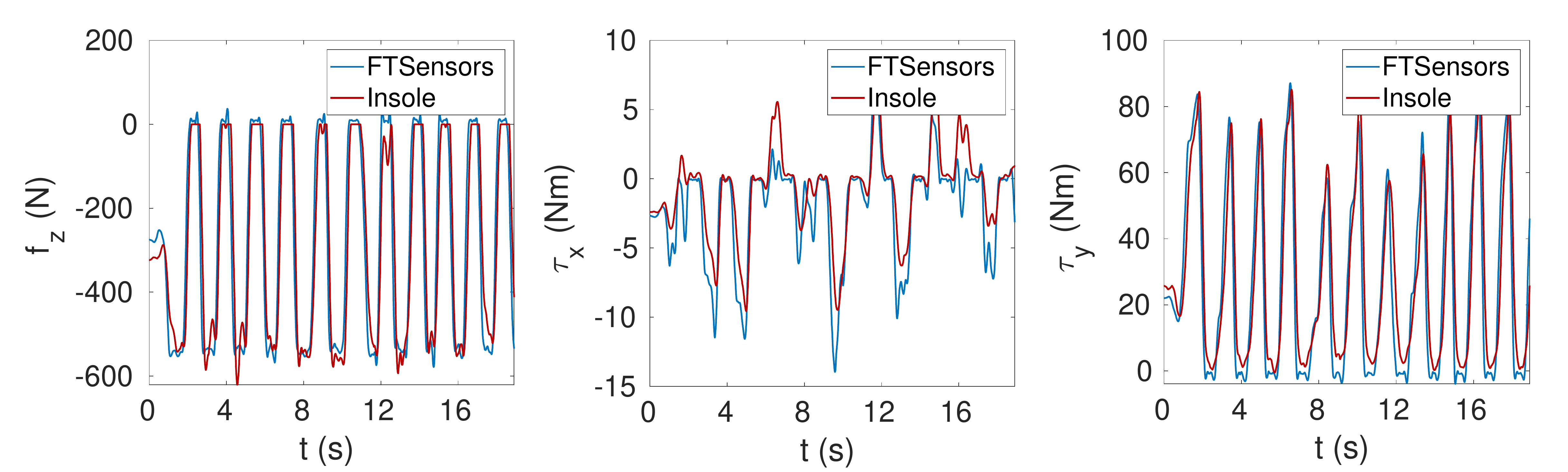}
		\caption{}
		\label{fig:walk_1_tracking}
	\end{subfigure}
	
	\caption{Vertical force $f_z$ and horizontal moments $m_x$ and $m_y$ tracking. ({\bf{a}}) Task T3. ({\bf{b}}) Task T4.}
	\label{fig:tracking}
\end{figure}

The validation process involved the comparison between the data measured by the FT sensors mounted on the sensorised shoes (in which the insoles were placed) and the data estimated with the calibrated insoles starting from the measured capacitances and using the relations (\ref{force_z_sk}) and (\ref{moments_sk}). The validation was performed in MATLAB. The data sets for the validation phase were not filtered to discard capacitances with high variation, as done with the calibration data sets. Table \ref{RMSE_validation} lists the maximum pressures sensed by the taxels in each task. It is worth specifying that the pressure was not uniformly distributed, and thus only some taxels sensed the maximum value. The accuracy of the calibration was evaluated for different tasks through the estimation of the root mean square error (RMSE) resulting from that comparison (Table \ref{RMSE_validation}). The RMSE varied from 7 to 95 N for the vertical force $f_z$, from $0.7$ to $1~\si{\newton\meter}$ for the moment about the $x$ axis $m_x$ and from 2  to $8~\si{\newton\meter}$ for the moment about the $y$ axis. The RMSE computed for tasks T1 and T2 was obtained at steady state; indeed, the experiments consisted of staying still on the insoles without changing the foot position. We can see that the RMSE increased for tasks  T3 to T6. The obtained results are justified by the introduction of dynamic movements on the insole, where the dynamics was faster for tasks from T4 to T6, which were the tasks that reached the largest error. In addition to the dynamics that involved transition phases, we needed to consider another factor influencing the results. Indeed, the insole was subjected to a different pressure intensity and pressure distribution than the static case. Tasks from T3 to T6 involved a greater vertical force on smaller areas corresponding to the forefoot and the heel. To overcome this problem, one solution could be the introduction of a dynamic sensors model in place of the static one adopted in this study. In conclusion, it is important to highlight that performances related to the steady state and the transition phase were different. For a better interpretation of the results, Table \ref{Error_percentage} shows the  tracking accuracy of the variables $f_z$, $m_x$ and $m_y$ \cite{yaramasu2016model}. As already commented, the accuracy was better for the static tasks.
In Figure \ref{fig:tracking}, we can see the estimated force and moments for the Task T3 and Task T4 data sets. The graph indicates that the estimated force $f_z$ and moments $m_x$, $m_y$  closely followed the actual force and moments applied on the insole.

\begin{table}[h]
	\setlength{\tabcolsep}{12pt}
\centering 
\caption{{Root Mean Square Error (RMSE) analysis}%caption has been moved to here please confirm and change table into three wires table. OK for the caption
~of vertical ground reaction force $f_z~(\si{\newton})$ and horizontal moments $m_x$, $m_y$ $(\si{\newton\meter})$, respectively, from tasks T1 to T6.}
 \label{RMSE_validation}
\begin{tabular}{c l c c c c}
\toprule
\textbf{Task} & \textbf{Description} & \thead{\textbf{Highest Pressure} \\ $(\si{\kilo\pascal})$} & \thead{\textbf{RMSE} \\ $f_z~(\si{\newton})$} & \thead{\textbf{RMSE} \\ $m_x~(\si{\newton\meter})$} & \thead{\textbf{RMSE} \\ $m_y~(\si{\newton\meter})$} \\
\midrule
T1 & Still on feet & 220 & 7.5946 & 0.70729 & 2.2892 \\

T2 & Still on left foot & 560 & 11.9475 & 1.8784 & 4.5675 \\

T3 & Slow movements on feet & 490 & 34.1682 & 1.8559 & 5.2909 \\

T4 & Walk 1 & 630 & 93.4157 & 2.1891 & 12.129 \\

T5 & Walk 2 & 600 & 84.8782 & 1.9286 & 11.0665 \\

T6 & Walk 3 & 360 & 95.1694 & 0.99215 & 7.7857 \\
\bottomrule
\end{tabular}

\end{table}

\begin{table}[h]
	\setlength{\tabcolsep}{12pt}
\centering 
\caption{{Tracking accuracy}%caption has been moved to here please confirm and change table into three wires table. OK for the caption
~of vertical ground reaction force $f_z$ and horizontal moments $m_x$, $m_y$, 
 respectively, from tasks T1 to T6.}
 \label{Error_percentage}
\begin{tabular}{c l c c c}
\toprule
\textbf{Task} & \textbf{Description} & \thead{\textbf{Accuracy (\%)} \\ $f_z$} & \thead{\textbf{Accuracy (\%)} \\ $m_x$} & \thead{\textbf{Accuracy (\%)} \\ $m_y$} \\
\midrule
T1 & Still on feet & 98.0472 & 70.8789 & 91.1336 \\

T2 & Still on left foot & 98.4868 & 73.6694 & 91.4024 \\

T3 & Slow movements on feet & 89.3013 & 71.6058 & 83.9142 \\

T4 & Walk 1 & 81.8419 & 64.056 & 75.9248 \\

T5 & Walk 2 & 83.8964 & 71.886 & 81.69 \\

T6 & Walk 3 & 82.5655 & 63.0632 & 70.7365 \\
\bottomrule
\end{tabular}

\end{table}

\begin{figure}[H]
\centering
	\includegraphics[width=\columnwidth]{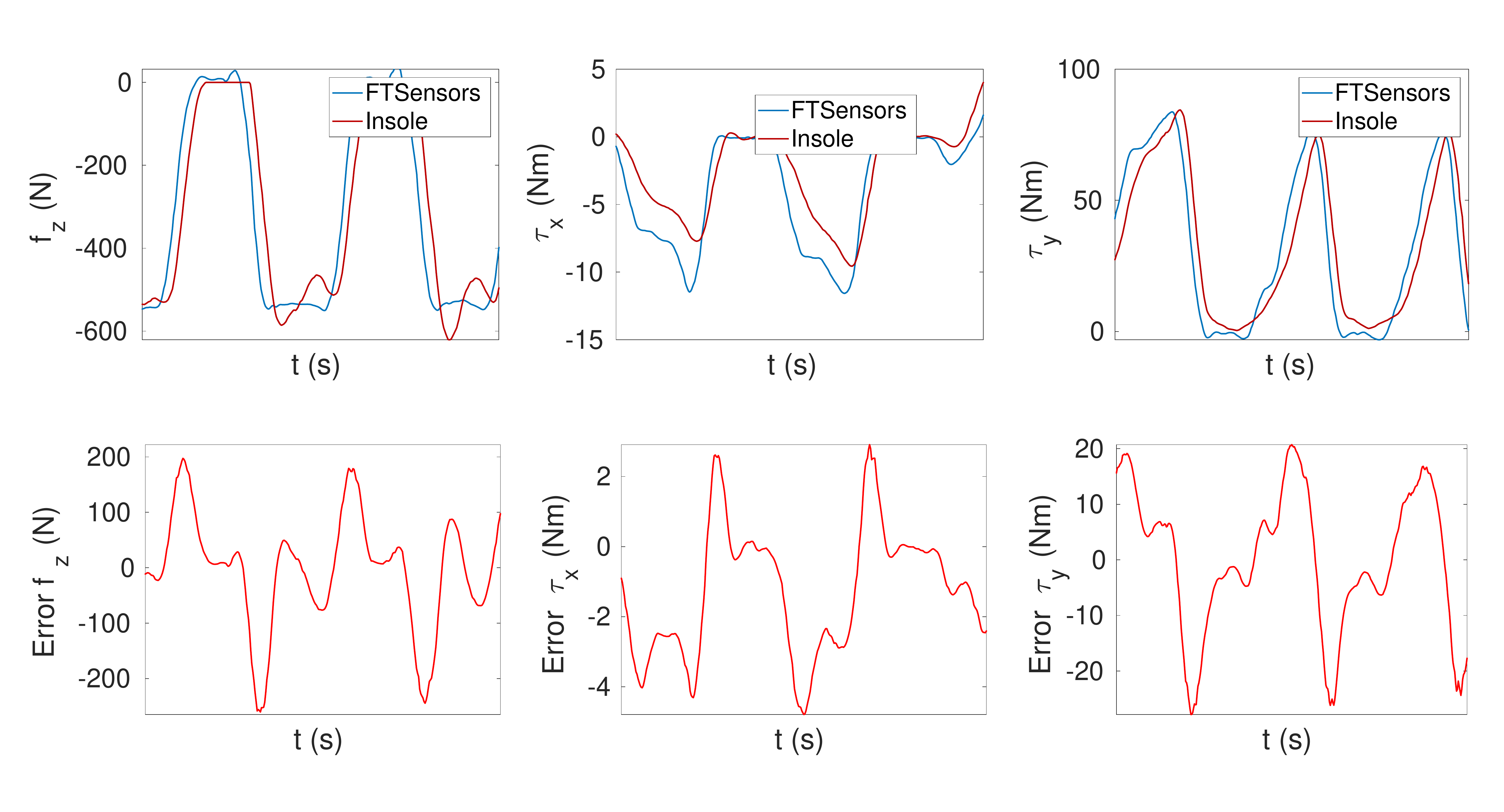}  
	\caption{Zoom{ed view of the}%was "zoom  on", confirm that your intended meaning is retained
~tracking error to  highlight the delay between FT sensors and insole estimations.}
	\label{fig:Error_Tracking_Missmatch}
\end{figure}

Figure \ref{fig:Error_Tracking_Missmatch} shows a delay of the insole estimation with respect to the FT sensor measurements (samples of FT sensor data sets and insole data sets were synchronised over time based  on the  receiver timestamp). This delay caused a very large error during the transition phase of up to 200 N. 

%Figure \ref{fig:Error_Tracking_Missmatch} highlights that the estimation is more accurate when the pressure variation over the insole is not fast. At steady state the error is very small. Indeed, Figure \ref{fig:Error_Tracking_Missmatch} shows a delay of the insole estimation with respect to the FT sensors measurements (samples of FT sensors data sets and insole data sets are time synchronised).

Furthermore, the calibration goodness could be analysed by comparing the CoP estimated by the FT sensors and the insole. The CoP estimated by the insole was computed as:

\begin{equation}
CoP_x = \displaystyle \frac{\displaystyle \sum_{i=1}^{280} P(C_i) x_i}{\displaystyle  \sum_{i=1}^{280} P(C_i)} \; , \quad CoP_y = \displaystyle \frac{\displaystyle \sum_{i=1}^{280} P(C_i) y_i}{\displaystyle  \sum_{i=1}^{280} P(C_i)} \; ,
\end{equation}

\noindent where $x_i$ and $y_i$ represent the $x$ and $y$ components of the $i$-th taxel in the plane of the insole. Instead, the CoP estimated by the FT sensors was computed as:

\begin{equation}
CoP_x = \displaystyle - \frac {m_y} {f_{z}} \; , \quad CoP_y = \displaystyle  \frac {m_x} {f_{z}} \; .
\end{equation}

\begin{figure}[H]
	\centering
	\begin{subfigure}[b]{1\textwidth}
		\includegraphics[width=1\linewidth]{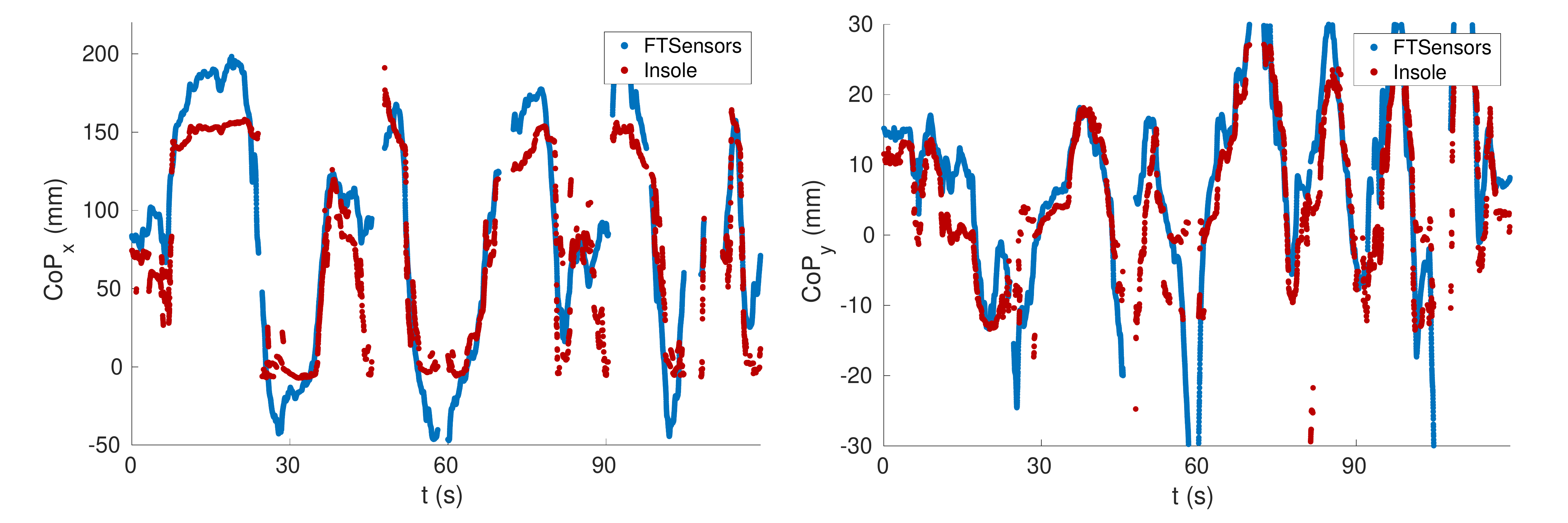}
		\caption{}
		\label{fig:CoP_slow_left_0_Ines} 
	\end{subfigure}
	
	\begin{subfigure}[b]{1\textwidth}
		\includegraphics[width=1\linewidth]{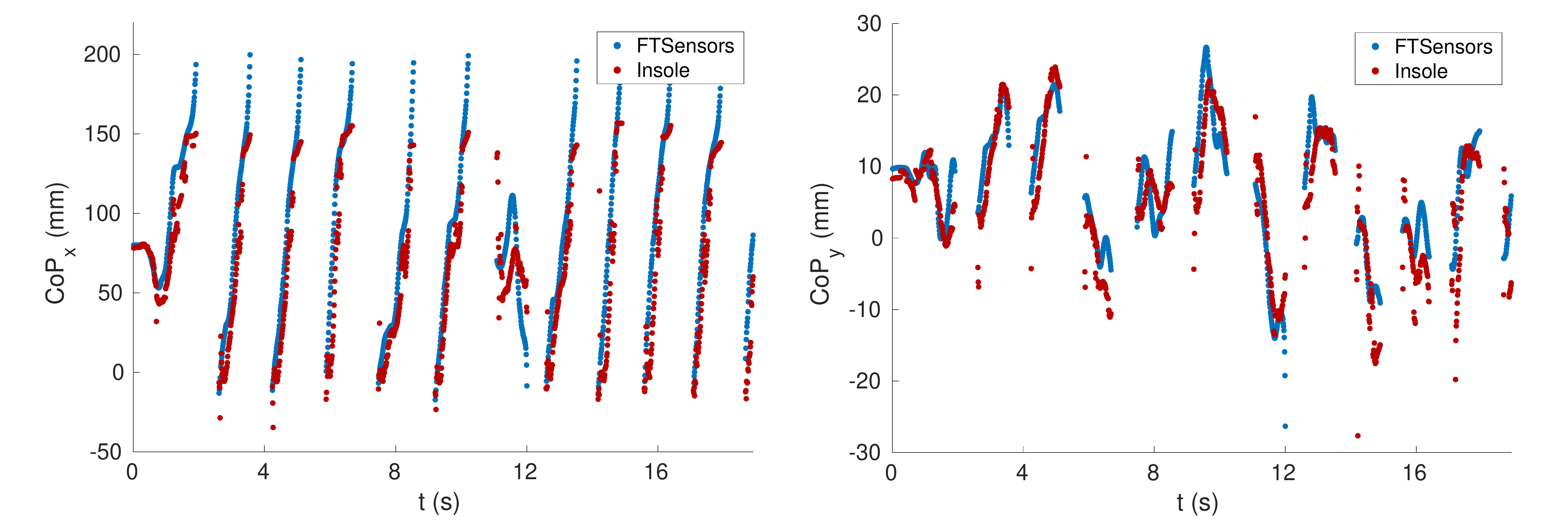}
		\caption{}
		\label{fig:CoP_Walk_1_Ines}
	\end{subfigure}
	
	\caption{Center of pressure (CoP) tracking. ({\bf{a}}) Task T3. ({\bf{b}}) Task T4.}
	\label{fig:tracking_CoP}
\end{figure}

Figure \ref{fig:tracking_CoP} shows the components $x$ and $y$ of the CoP related to Tasks T3 and T4. It can be observed that the values related to the insole were close to those  related to the FT sensors (i.e., the ground truth).

Table \ref{RMSE_CoP} shows the RMSE for several tasks. The estimation of the CoP varied between 11 and 38 mm for the $x$ component and between 3 and 7 mm for the $y$ component. Instead, Table \ref{percentage_error_cop} contains the tracking accuracy. It shows that, contrary to the force and moments tracking, the accuracy was higher for dynamic tasks.

\begin{table}[H]
	\setlength{\tabcolsep}{12pt}
	\centering 
\caption{{Root mean square error (RMSE) analysis}% Ok for the caption
		of centre of pressure $CoP_x$, $CoP_y$ (mm), respectively, from tasks T1 to T6.}
	\label{RMSE_CoP}
	\begin{tabular}{c l c c}
		\toprule
		\textbf{Task} & \textbf{Description} & \thead{\textbf{RMSE} \\ $CoP_x$ (mm)} & \thead{\textbf{RMSE} \\ $CoP_y$ (mm)} \\
		\midrule
		T1 & Still on feet & 18.6429 & 4.951 \\
		
		T2 & Still on left foot & 11.941 & 4.2918 \\
		
		T3 & Slow movements on feet & 28.9685 & 6.697 \\
		
		T4 & Walk 1 & 19.8642 & 3.8349 \\
		
		T5 & Walk 2 & 18.6675 & 3.1547 \\
		
		T6 & Walk 3 & 37.5296 & 5.084 \\
		\bottomrule
	\end{tabular}
	
\end{table}

\begin{table}
	\setlength{\tabcolsep}{12pt}
	\centering 
\caption{{Tracking accuracy of centre of pressure} $CoP_x$, $CoP_y$, respectively, from task T1 to T6.} % Ok for the caption
	\label{percentage_error_cop}
	\begin{tabular}{c l c c}
		\toprule
		\textbf{Task} & \textbf{Description} & \thead{\textbf{Accuracy [\%]} \\ $CoP_x$ } & \thead{\textbf{Accuracy [\%]} \\ $CoP_y$} \\
		\midrule
		T1 & Still on feet & 69.5057 & 89.1761 \\
		
		T2 & Still on left foot & 84.3097 & 60.71643 \\
		
		T3 & Slow movements on feet & 99.9442 & 99.9884 \\
		
		T4 & Walk 1 & 99.9805 & 99.9965 \\
		
		T5 & Walk 2 & 99.9811 & 99.997 \\
		
		T6 & Walk 3 & 99.9484 & 99.9935 \\
		\bottomrule
	\end{tabular}
	
\end{table}

\begin{figure}[H]
\centering
	\includegraphics[width=1\columnwidth]{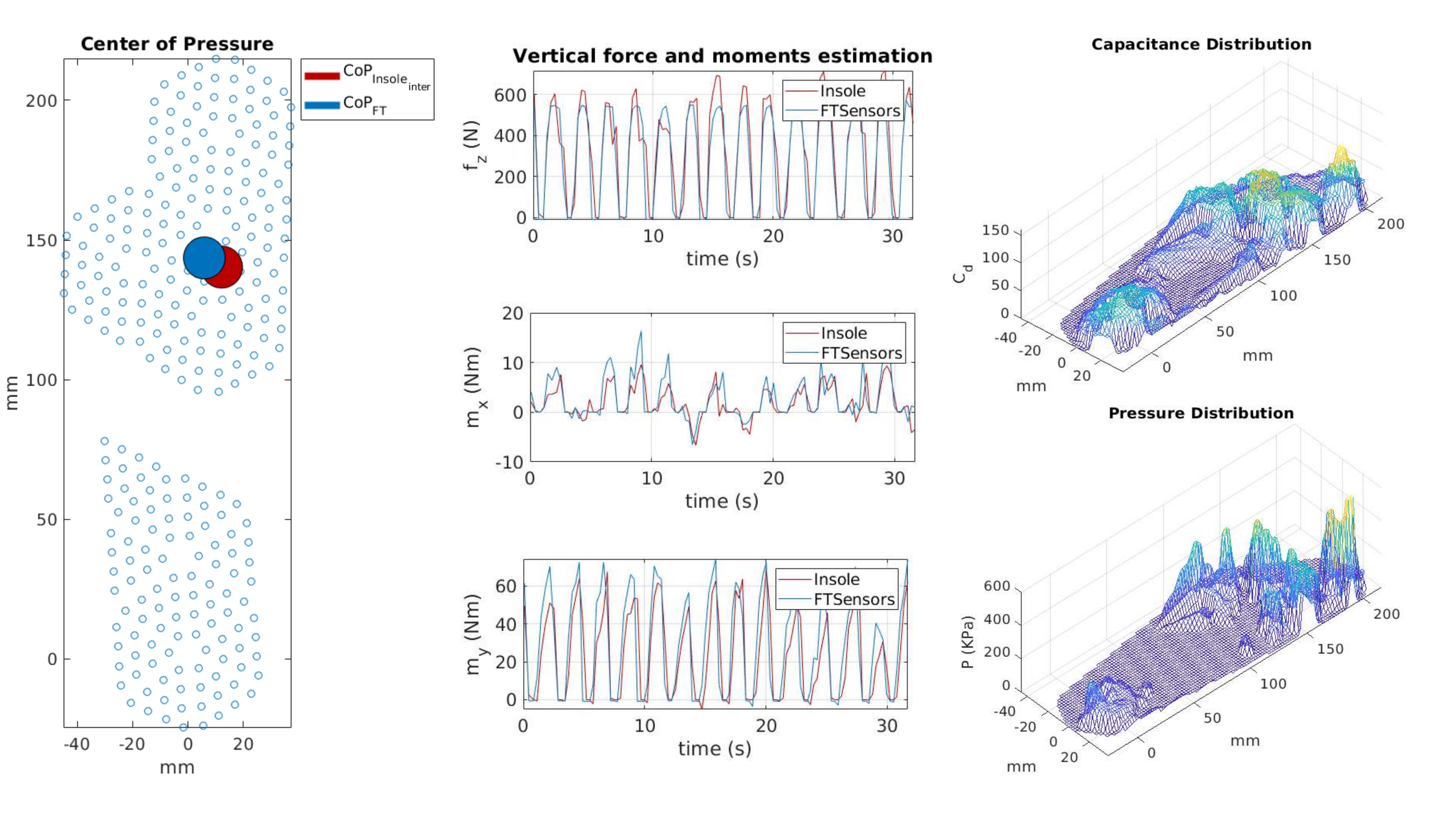}  
	\caption{Online visualisation tool for the insoles.  On the left-hand side%Can subfigure labels such as (a), (b), (c) be used?
, the comparison between centres of pressure estimated by force/torque (FT) sensors and insole. In the middle, estimation of vertical contact forces $f_z~(\si{\newton})$ and moments $m_x$, $m_y$~$(\si{\newton\meter})$ using the calibrated insole. On the right-hand side, the pressure distribution.}%Correct "Center" to "Centre" in the figure if appropriate
	\label{fig:matlab_online_visual}
\end{figure}

We developed a MATLAB-based tool for online visualisation of the estimated CoP, capacitances and pressure distribution, and estimated forces and moments as shown in Figure \ref{fig:matlab_online_visual}. Unfiltered data were streamed through the YARP protocol with a frequency of $50~\si{\hertz}$ and read in MATLAB by opening a YARP port \cite{metta_yarp:_2006}. Data were received and processed in real-time.  The cycle for reading, processing and  visualising the results had a frequency of $27~\si{\hertz}$. This means that we processed one sample for every two sent, where the sample indicates the set of taxel measurements in a given instant. 

On the left, the figure shows the comparison between the centre of pressure estimated from the FT sensor data and that estimated by the insole. In the middle, the graph contains the comparison between vertical contact force and horizontal moments estimated through the FT sensors and the insole, respectively. On the right, the capacitance distribution is given by the taxel measurements and the pressure distribution was computed using the coefficients estimated during the calibration phase. For visualisation purposes, the estimated forces and the pressure distribution are positive. The insole is very sensitive and thus the measured capacitance values were not equal to the rest value even when there was no weight on the foot. This is due to the contact between the foot and the insole, which becomes strong when the shoes are strapped. To avoid considering fake pressures, we used a threshold on the measured capacitance values set to 30. Furthermore, the pressure values were  interpolated over the entire insole surface in order to obtain a more accurate estimation of the pressure distribution and the CoP. Additionally, the capacitance values were interpolated to visualise the capacitance distribution. A video of the online visualisation tool is provided to show the results of the validation in real time.

%\iffalse
%\fi

%% file: sections/conclusions.tex
%!TEX root = ../IS_sensors2019.tex

\section{Conclusions}\label{conclusions}

% Conclusions
% -----------

In this paper, we presented a novel insole prototype for the real-time monitoring of plantar pressure distributions. Compared to the existing devices, the measurements are not affected by temperature changes because of an internal compensation. The sensors cover almost the entire surface of the insole, having a high spatial resolution. 
We defined and identified a model for each single taxel through a proved calibration procedure, which is required for an accurate estimation of the pressure distribution, vertical forces and moments about the horizontal axes. The insole needs to be calibrated {only once}%was "{una tantum}", confirm
~and provides a calibration model for each individual taxel. The choice of materials allows a cost-effective wearable sensor that is almost comparable in performance with other expensive solutions.
We showed that the calibration of the insole gave good results and the estimated variables were quite reliable when compared to those measured and estimated with the FT sensors. We also described an example of a real scenario, demonstrating that it was possible to obtain good results for real-time applications.

Nevertheless, the proposed prototype can still be improved. 
A wireless module can be used to remove cables. The PCB and the support for the electronics can be more stretchable and soft to provide a more comfortable wearing experience for the user.  
Additionally, we can improve the calibration with the following additional mathematical considerations:
\begin{itemize}
	\item A weighted iterative least-square method can be used in order to avoid manually tuning the weights;
	\item The formulation for describing the relationship between capacitance and pressure for each taxel could also take into account some hysteresis in the stiffness of the skin dielectric and the electric state of the sensor;
	\item Constraints on the model can be added in the optimisation problem;
	\item The accuracy and the repeatability of the data collection procedure can be improved.
\end{itemize}

As future work, we aim to redesign the calibration device mentioned in Section~\ref{subsec:calibration_joan} in order to reach higher pressures and control the pressure variation applied on the insole. Finally, we wish to keep improving the insole design with the desirable hardware characteristics mentioned above.